\UseRawInputEncoding
\documentclass[lettersize,journal]{IEEEtran}
\usepackage{placeins}
\usepackage{pgfplots}
\usepackage{amsmath, amsthm}
\usepackage[OT1]{fontenc}
\usepackage[english]{babel}

\usepackage{algorithmic}
\usepackage{array}
\usepackage{textcomp}
\usepackage{stfloats}
\usepackage{url}
\usepackage{verbatim}
\usepackage[nolist]{acronym}
\usepackage[normalem]{ulem}

\usepackage{nicefrac}
\usepackage{tikz}
\usetikzlibrary{fit}
\usepackage{graphicx}
\usepackage{xcolor}

\usepackage{tikz}
\usepackage{pgfplots, pgfplotstable}
\pgfplotsset{compat=1.18}
\usepackage{amssymb}
\usepackage{subfigure}
\usepackage{cite}
\usepackage{dsfont}

\usepackage{threeparttable}

\usepackage{stfloats}  % to keep fig near the text
\usepackage[linesnumbered,ruled,vlined]{algorithm2e}

\usepackage{etoolbox}
\makeatletter
\patchcmd{\@algocf@pre@ruled}{\hrule}{\vspace*{5pt}\hrule}{}{}
\makeatother

\usepackage{makecell}
\usepackage{booktabs}

\usepackage{arydshln}
\newcolumntype{C}[1]{>{\centering\let\newline\\\arraybackslash\hspace{0pt}}m{#1}}
\newcolumntype{L}{>{\arraybackslash}p{0.7\linewidth}}
\newcolumntype{N}{@{}m{0pt}@{}}

\usepgfplotslibrary{statistics,groupplots}
\usetikzlibrary{dsp,calc,plotmarks}

\usepackage{multirow}
\usepackage{standalone}
\usepackage{balance}
\usepackage[capitalise]{cleveref}
\usepgfplotslibrary{groupplots}

\Crefformat{figure}{#2Figure~#1#3}
\Crefmultiformat{figure}{Figures~#2#1#3}{ and~#2#1#3}{, #2#1#3}{ and~#2#1#3}

\DeclareMathOperator*{\argmin}{arg\,min}

\usetikzlibrary{calc}

\pgfplotsset{
    legend image with text/.style={
        legend image code/.code={%
            \node[anchor=center] at (0.3cm,0cm) {#1};
        }
    },
    label style = {font=\fontsize{9pt}{7.2}\selectfont},
    tick label style = {font=\fontsize{7pt}{7.2}\selectfont}
}

\definecolor{matplottikz-color1}{HTML}{144b32}
\definecolor{matplottikz-color2}{HTML}{db7622}
\definecolor{matplottikz-color3}{HTML}{314c82}
\definecolor{matplottikz-color4}{HTML}{c71a1f}
\definecolor{matplottikz-color5}{HTML}{1e73af}
\definecolor{matplottikz-color6}{HTML}{bf8fa2}
\definecolor{matplottikz-color7}{HTML}{d76672}
\definecolor{matplottikz-color8}{HTML}{7a6d49}
\definecolor{matplottikz-color9}{HTML}{318d43}
\definecolor{matplottikz-color10}{HTML}{28b5a3}

\pgfplotsset{
    compat=1.18,
    every axis plot/.append style={
        line width=0.9pt,
        mark size=1.8pt
    },
    every axis/.append style={
        label style={
            font=\fontsize{9pt}{7.2pt}\selectfont
        },
        tick label style={
            font=\fontsize{7pt}{7.2pt}\selectfont
        },
        grid=major,
        grid style={dashed,gray!40}
    }
}

\pgfplotsset{
    fssc/.style={
         thick,
        red!85!black,
        dotted,
        mark=*,
        mark options={solid,fill=red!85!black, draw=red!85!black}
    }
}

\pgfplotsset{
    spae8/.style={
        thick,
        blue!70!black,
        solid,
        mark=square*,
        mark options={
            solid,
            fill=white,
            draw=blue!70!black,
            line width=0.6pt
        }
    }
}

\pgfplotsset{
    spae16/.style={
        thick,
        teal!70!black,
        solid,
        mark=triangle*,
        mark options={
            solid,
            fill=white,
            draw=teal!70!black,
            line width=0.6pt
        }
    }
}

\pgfplotsset{
    spae32/.style={
        thick,
        green!60!black,
        solid,
        mark=diamond*,
        mark options={
            solid,
            fill=white,
            draw=green!60!black,
            line width=0.6pt
        }
    }
}

\pgfplotsset{
    scl8/.style={
        thick, magenta!80!black, dotted, mark=pentagon*, mark options={solid,fill=magenta!80!black,draw=magenta!80!black,line width=0.6pt}
    }
}

\pgfplotsset{
    ldpc5/.style={
        thick,
        orange!90!black,
        dash pattern=on 4pt off 1.5pt,
        mark=star,
        mark options={
            solid,
            fill=white,
            draw=orange!90!black,
            line width=0.6pt
        }
    }
}

\pgfplotsset{
    ldpc10/.style={
        thick,
        brown!80!black,
        dash pattern=on 3pt off 2pt on 1pt off 2pt,
        mark=star,
        mark options={
            solid,
            fill=white,
            draw=brown!80!black,
            line width=0.6pt
        }
    }
}

\pgfplotsset{
    ldpc20/.style={
        thick,
        violet!80!black,
        dash pattern=on 2pt off 1pt on 1pt off 1pt,
        mark=star,
        mark options={
            solid,
            fill=white,
            draw=violet!80!black,
            line width=0.6pt
        }
    }
}

\pgfplotsset{
    aed8/.style={
     thick, 
     cyan!70!black, 
     dotted, 
     mark=halfcircle, 
     mark options={solid,fill=white,draw=cyan!70!black,line width=0.6pt}
    }
}

\pgfplotsset{
    construction-sc/.style={
        thick,
        black!70,
        dashed,
        mark=*,
        mark options={solid,fill=white,draw=black!70,line width=0.6pt
        }
    }
}

\pgfplotsset{
    construction-scl8/.style={
        thick,
        black!70,
        dashdotted,
        mark=pentagon*,
        mark options={solid,fill=white,draw=black!70,line width=0.6pt}
    }
}

\pgfplotsset{
    ppv/.style={
        black,
        solid,
        line width=1.2pt,
        no markers
    }
}

\pgfplotsset{
    5g-scl8-crc/.style={
        thick,
        red!85!black,
        dashdotdotted,
        mark=square*,
        mark options={solid,fill=red!85!black,draw=red!85!black,line width=0.6pt}
    }
}

\providecommand{\PolarDecoderLegend}{%
    \begingroup
    \footnotesize
    \setlength{\fboxsep}{4pt}%
    \setlength{\tabcolsep}{5pt}%
    \renewcommand{\arraystretch}{1.25}%
    \fbox{%
        \begin{tabular}{@{}l@{\hspace{6pt}}c@{\hspace{6pt}}l@{}}

            \begin{tabular}[c]{@{}l@{}}
                Symmetric Polar \\
                5G Polar        \\
                5G LDPC         \\
                Bound
            \end{tabular}

            &

        \rule[-4.9ex]{0.4pt}{10.8ex}
            &
            \begin{tabular}[c]{@{}llllll@{}}
                \ref{leg:sym-sc}~SC
                    & \ref{leg:sym-scl}~SCL-8
                    & \ref{leg:ae-sc}~AE-SC-8
                    & \ref{leg:aedisc-8}~AE-DISC-8
                    & \ref{leg:aedisc-16}~AE-DISC-16
                    & \ref{leg:aedisc-32}~AE-DISC-32
                    \\
                \ref{leg:5g-sc}~SC
                    & \ref{leg:5g-scl}~SCL-8
                    &\ref{leg:scl8_crc}~SCL-8-CRC11
                    & & &
                    \\
                \ref{leg:ldpc-5}~LMS-5
                    & \ref{leg:ldpc-10}~LMS-10
                    & \ref{leg:ldpc-20}~LMS-20
                    & & &
                    \\
                \multicolumn{3}{@{}l}{\ref{leg:ppv}~PPV Meta Converse}  & & &
            \end{tabular}

        \end{tabular}%
    }%
    \endgroup
}

\providecommand{\RMDecoderLegend}{%
    \begingroup
    \footnotesize
    \setlength{\fboxsep}{4pt}%
    \setlength{\tabcolsep}{5pt}%
    \renewcommand{\arraystretch}{1.25}%
    \fbox{%
        \begin{tabular}{@{}l@{\hspace{6pt}}c@{\hspace{6pt}}l@{}}

            \begin{tabular}[c]{@{}l@{}}
                RM              \\
                5G Polar        \\
                5G LDPC         \\
                Bound
            \end{tabular}

            &

        \rule[-4.9ex]{0.4pt}{10.8ex}
            &
            \begin{tabular}[c]{@{}llllll@{}}
                \ref{leg:sym-sc}~SC
                    & \ref{leg:sym-scl}~SCL-8
                    & \ref{leg:ae-sc}~AE-SC-8
                    & \ref{leg:aedisc-8}~AE-DISC-8
                    & \ref{leg:aedisc-16}~AE-DISC-16
                    & \ref{leg:aedisc-32}~AE-DISC-32
                    \\
                \ref{leg:5g-sc}~SC
                    & \ref{leg:5g-scl}~SCL-8
                    & \ref{leg:scl8_crc}~SCL-8-CRC11& & &
                    \\
                \ref{leg:ldpc-5}~LMS-5
                    & \ref{leg:ldpc-10}~LMS-10
                    & \ref{leg:ldpc-20}~LMS-20
                    & & &
                    \\
                    \multicolumn{3}{@{}l}{\ref{leg:ppv}~PPV Meta Converse}  & & &
            \end{tabular}

        \end{tabular}%
    }%
    \endgroup
}

\providecommand{\MismatchRateLegend}{%
    \begingroup
    \footnotesize
    \setlength{\fboxsep}{4pt}%
    \setlength{\tabcolsep}{5pt}%
    \renewcommand{\arraystretch}{1.25}%
    \fbox{%
        \begin{tabular}{@{}lll@{}}
            \ref{leg:mismatch-aedisc-8}~AE-DISC-8
            &
            \ref{leg:mismatch-aedisc-16}~AE-DISC-16
            &
            \ref{leg:mismatch-aedisc-32}~AE-DISC-32
        \end{tabular}%
    }%
    \endgroup
}

\pgfplotsset{
    AEDFP128/.style={
        thick,
        red!85!black,
        solid,
        mark=*,
        mark options={
            solid,
            fill=none,
            draw=red!85!black
        }
    },
    AEDQ128/.style={
        thick,
        blue!70!black,
        dashed,
        mark=square*,
        mark options={
            solid,
            fill=white,
            draw=blue!70!black,
            line width=0.6pt
        }
    },
    AEDFP256/.style={
        thick,
        red!60!black,
        solid,
        mark=triangle*,
        mark options={
            solid,
            fill=none,
            draw=red!60!black
        }
    },
    AEDQ256/.style={
        thick,
        blue!50!black,
        dashed,
        mark=diamond*,
        mark options={
            solid,
            fill=white,
            draw=blue!50!black,
            line width=0.6pt
        }
    }
}

\definecolor{matplottikz-color1}{HTML}{144b32}
\definecolor{matplottikz-color2}{HTML}{db7622}
\definecolor{matplottikz-color3}{HTML}{314c82}
\definecolor{matplottikz-color4}{HTML}{c71a1f}
\definecolor{matplottikz-color5}{HTML}{1e73af}
\definecolor{matplottikz-color6}{HTML}{bf8fa2}
\definecolor{matplottikz-color7}{HTML}{d76672}
\definecolor{matplottikz-color8}{HTML}{7a6d49}
\definecolor{matplottikz-color9}{HTML}{318d43}
\definecolor{matplottikz-color10}{HTML}{28b5a3}

\title{AE-DISC: Automorphism Ensemble Direct-Invocation Successive Cancellation for Low-Latency Decoding} 

\author{Sihui Shen, Ryan Seah, Marvin R\"ubenacke, and Warren J. Gross \\
Department of Electrical and Computer Engineering, McGill University, Montr\'eal, Qu\'ebec, Canada \\
Emails: \{sihui.shen, ryan.seah\}@mail.mcgill.ca, \{marvin.ruebenacke, warren.gross\}@mcgill.ca
\thanks{Parts of this work have been submitted for presentation at IEEE ICC 2027~\cite{shen2026lowlatency}.}}

\begin{document}
\begin{acronym}
    \acro{AED}{automorphism ensemble decoding}
    \acro{AR/VR}{augmented reality/virtual reality}
    \acro{ASIC}{application-specific integrated circuit}
    \acro{AWGN}{additive white Gaussian noise}
    \acro{BER}{bit error rate}
    \acro{BLTA}{block lower-triangular affine}
    \acro{BP}{belief propagation}   
    \acro{BPSK}{binary phase shift keying}
    \acro{CRC}{cyclic redundancy check}
    \acro{FER}{frame error rate}
    \acro{FSSC}{fast simplified successive cancellation}
    \acro{FHT}{fast Hadamard transform}
    \acro{HRLLC}{hyper-reliable low-latency communication}
    \acro{URLLC}{Ultra-Reliable Low-Latency Communication}
    \acro{IoT}{internet of things}
    \acro{LDPC}{low-density parity-check}
    \acro{LLR}{log-likelihood ratio}
    \acro{LMS}{layered min-sum}
    \acro{NR}{New Radio}
    \acro{PE}{processing element}
    \acro{PPV}{Polyanski--Poor--Verd\'u}
    \acro{REP}{Repetition}
    \acro{RM}{Reed--Muller}
    \acro{R0}{Rate-0}
    \acro{R1}{Rate-1}
    \acro{SC}{successive cancellation}
    \acro{SCL}{successive cancellation list}
    \acro{FSSCL}{fast simplified successive cancellation list}
    \acro{SPC}{single parity-check}
    \acro{SNR}{signal-to-noise ratio}
    \acro{SPAE}{speculative successive cancellation automorphism ensemble}
    \acro{UPO}{universal partial order}
    \acro{GMC}{generalized multiple concatenated}
    \acro{Spec-SC}{speculative successive cancellation}
    \acro{DISC}{direct invocation successive cancellation}
    \acro{AE-DISC}{automorphism ensemble direct-invocation successive cancellation}
    \acro{AE-SC}{automorphism ensemble successive cancellation}
    \acro{PM}{path metric}
\end{acronym}

\maketitle

\begin{abstract}
Emerging applications in \ac{HRLLC} demand channel decoders that deliver deterministic performance under stringent latency and reliability constraints.
While Polar codes are well suited for this scenario due to their strong error-correction capability,  conventional \ac{SC} decoding exhibits inherent serial dependencies that lead to high decoding latency.
This paper presents the \ac{AE-DISC} decoder, a low-latency framework that enables the concurrent computation of the left and right decoding subtrees in the \ac{SC} process, yielding substantial speed-ups.
By exploiting permutation automorphisms, \ac{AE-DISC} decoding further improves error-correction performance. %
Simulation results demonstrate that \ac{AE-DISC} attains up to 1.4\,dB gain over the 5G \ac{SC} baseline and surpasses \ac{LMS} \ac{LDPC} decoders at a \ac{FER} of $10^{-6}$, while maintaining fixed algorithmic complexity. 
Both theoretical analysis and \ac{ASIC} synthesis results further show up to 2 times decoding speed-up relative to \ac{FSSC}. 
\end{abstract}
\acresetall


\section{Introduction}

\IEEEPARstart{M}{odern} wireless communication systems, such as 5G \ac{NR}, employ channel coding schemes tailored to different channel types to meet stringent performance requirements~\cite{5g_coding_overview}. \Ac{LDPC}~\cite{LDPC_original, mackay1996near, Ryan2005AnIT} codes are used in data channels for their execellent error-correction performance at long block length, and high-throughput transmission enabled by parallel decoding. 
In contrast, Polar codes~\cite{polar_codes_arikan} are employed in the control channels, where high reliability and low decoding latency are critical. The high reliability requirements are commonly addressed using \ac{SCL} decoding at the cost of increased decoding latency compared with plain \ac{SC} decoding~
\cite{sarkis2014fast, sarkis2015fast, tal2015list,fan2015low,hashemi2017fast,alamdar2011simplified,balatsoukas2015llr,cocskun2022information}.

As wireless systems evolve toward 6G and beyond, there is growing demand for \ac{HRLLC}~\cite{pradhan2024survey, hrllc_proof_concept}. \Ac{HRLLC} supports mission-critical applications that require both extreme reliability, packet error rates as low as \(10^{-5}\) to \(10^{-7}\), and ultra-low latency on the order of 1\,ms or less~\cite{3gppR1_2507061_huawei}. Applications include autonomous driving, industrial automation, and the tactile internet~\cite{3GPP_TS_22_870, 3GPP_TS_22_104, shirvanimoghaddam2018short, ibrahim2025urllc}. 
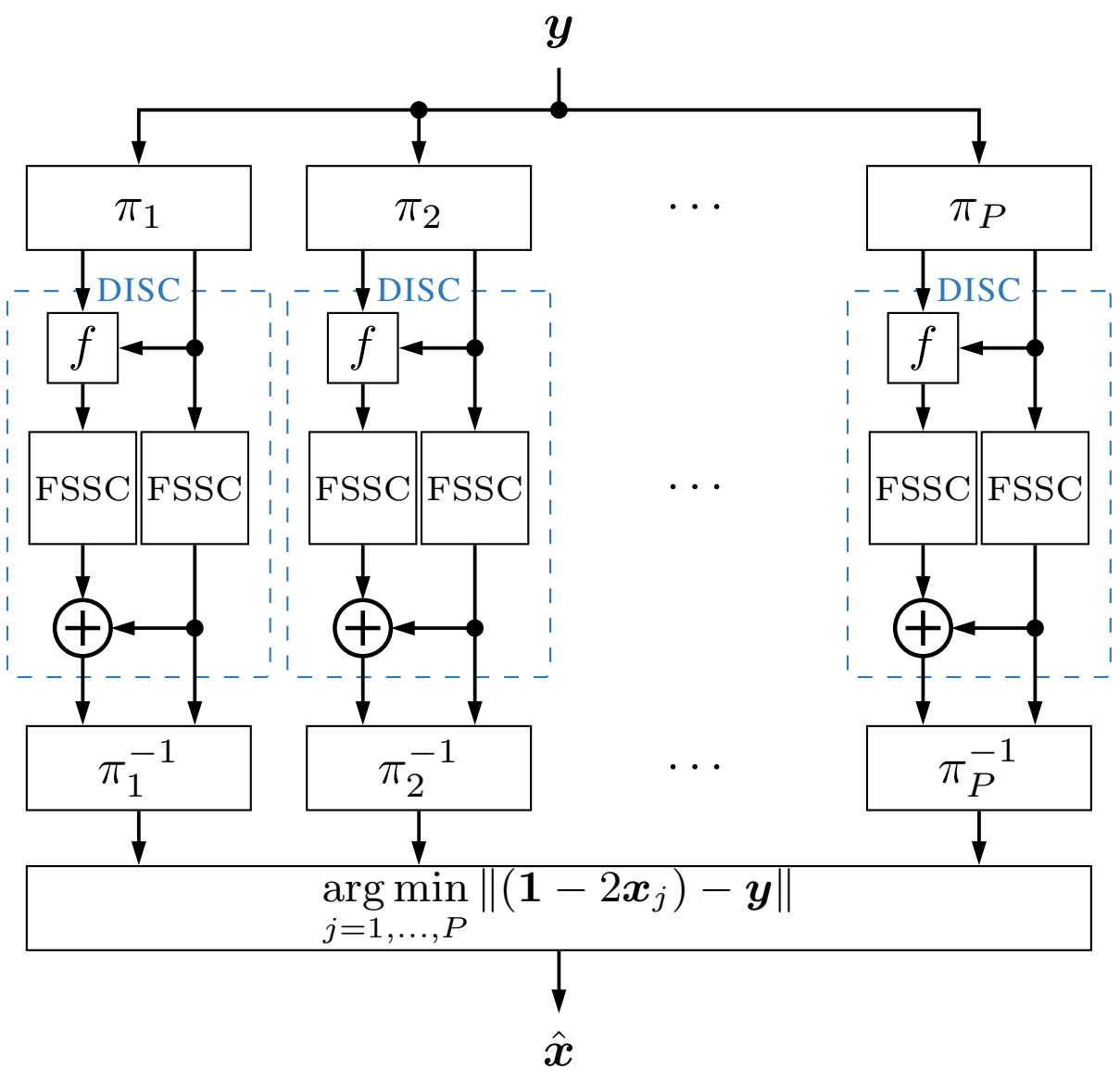
\begin{figure}[t]
    \centering
    \definecolor{discblue}{RGB}{47,121,185}

\begin{tikzpicture}
    \tikzset{
        decrect/.style={
            rectangle,
            draw,
            minimum width=.7cm,
            minimum height=.8cm,
            fill=white,
            inner sep=1pt,
            font=\scriptsize,
            align=center
        },
        intrect/.style={
            rectangle,
            draw,
            minimum width=1.6cm,
            minimum height=.6cm,
            fill=white,
            inner sep=0pt
        },
        frect/.style={
            rectangle,
            draw,
            minimum size=.5cm,
            fill=white,
            inner sep=0pt
        },
        decision/.style={
            rectangle,
            draw,
            minimum width=7.6cm,
            minimum height=.6cm,
            fill=white,
            inner sep=0pt,
            font=\small
        }
    }

    % Received vector
    \node (Lch) at (0,.3) {$\boldsymbol{y}\mathstrut$};

    % Distribution nodes
    \node[dspnodefull] (s1) at (-1,-.3) {};
    \node[dspnodefull] (s2) at (0,-.3) {};

    % Automorphisms
    \node[intrect] (int1) at (-3,-1)
        {$\pi_{1}\mathstrut$};
    \node[intrect] (int2) at (-1,-1)
        {$\pi_{2}\mathstrut$};
    \node at (1,-1) {$\cdots$};
    \node[intrect] (int3) at (3,-1)
        {$\pi_{P}\mathstrut$};

    % Root f-functions and direct speculative branches
    \node[frect] (f1) at (-3.4,-2) {$f$};
    \node[dspnodefull] (dot1) at (-2.6,-2) {};

    \node[frect] (f2) at (-1.4,-2) {$f$};
    \node[dspnodefull] (dot2) at (-0.6,-2) {};

    \node[frect] (f3) at (2.6,-2) {$f$};
    \node[dspnodefull] (dot3) at (3.4,-2) {};

    % Parallel component decoders:
    % SC for polar codes and GMC for RM codes
    \node[decrect] (da1) at (-3.4,-3)
        {\shortstack{\(\mathrm{FSSC}\)}};
    \node[decrect] (db1) at (-2.6,-3)
        {\shortstack{\(\mathrm{FSSC}\)}};

    \node[decrect] (da2) at (-1.4,-3)
        {\shortstack{\(\mathrm{FSSC}\)}};
    \node[decrect] (db2) at (-0.6,-3)
        {\shortstack{\(\mathrm{FSSC}\)}};

    \node at (1,-3) {$\cdots$};

    \node[decrect] (da3) at (2.6,-3)
        {\shortstack{\(\mathrm{FSSC}\)}};
    \node[decrect] (db3) at (3.4,-3)
        {\shortstack{\(\mathrm{FSSC}\)}};

    % Codeword reconstruction
    \node[dspadder] (x1) at (-3.4,-4) {};
    \node[dspnodefull] (dotb1) at (-2.6,-4) {};

    \node[dspadder] (x2) at (-1.4,-4) {};
    \node[dspnodefull] (dotb2) at (-0.6,-4) {};

    \node[dspadder] (x3) at (2.6,-4) {};
    \node[dspnodefull] (dotb3) at (3.4,-4) {};

% =====================================================
% Dashed box around each decoding path
% =====================================================
\node[
    draw=discblue,
    dashed,
    rounded corners=1pt,
    inner xsep=1.5mm,
    inner ysep=1.5mm,
    fit=(f1)(dot1)(da1)(db1)(x1)(dotb1)
](disc1) {};

\node[
    draw=discblue,
    dashed,
    rounded corners=1pt,
    inner xsep=1.5mm,
    inner ysep=1.5mm,
    fit=(f2)(dot2)(da2)(db2)(x2)(dotb2)
](disc2) {};

\node[
    draw=discblue,
    dashed,
    rounded corners=1pt,
    inner xsep=1.5mm,
    inner ysep=1.5mm,
    fit=(f3)(dot3)(da3)(db3)(x3)(dotb3)
](disc3) {};

% DISC labels
\node[
    font=\scriptsize,
    text=discblue,
    fill=white,
    inner xsep=2pt,
    inner ysep=0pt
] at (disc1.north) {\textsc{DISC}};

\node[
    font=\scriptsize,
    text=discblue,
    fill=white,
    inner xsep=2pt,
    inner ysep=0pt
] at (disc2.north) {\textsc{DISC}};

\node[
    font=\scriptsize,
    text=discblue,
    fill=white,
    inner xsep=2pt,
    inner ysep=0pt
] at (disc3.north) {\textsc{DISC}};

    % Inverse automorphisms
    \node[intrect] (dint1) at (-3,-5)
        {$\pi_{1}^{-1}\mathstrut$};
    \node[intrect] (dint2) at (-1,-5)
        {$\pi_{2}^{-1}\mathstrut$};
    \node at (1,-5) {$\cdots$};
    \node[intrect] (dint3) at (3,-5)
        {$\pi_{P}^{-1}\mathstrut$};

    % Candidate selection
    \node[decision] (d) at (0,-6)
        {$\displaystyle
        \argmin_{j=1,\ldots,P}
        \left\|
        (\boldsymbol{1}-2\boldsymbol{x}_{j})
        -\boldsymbol{y}
        \right\|$};

    % Final output
    \node (end) at (0,-7)
        {$\hat{\boldsymbol{x}}$};

    % Input connections
    \draw[dspconn] (Lch) -- (s2) -- (s1) -| (int1);
    \draw[dspconn] (s1) -- (int2);
    \draw[dspconn] (s2) -| (int3);

    % Repeated ensemble-path connections
    \foreach \i in {1,...,3} {
        % Permuted vector to the two parallel branches
        \draw[dspconn]
            (f\i.north |- int\i.south) -- (f\i);

        \draw[dspconn]
            (dot\i.north |- int\i.south)
            -- (dot\i)
            -- (f\i);

        % Component decoders
        \draw[dspconn] (f\i) -- (da\i);
        \draw[dspconn] (dot\i) -- (db\i);

        % Reconstruction
        \draw[dspconn] (da\i) -- (x\i);
        \draw[dspconn]
            (db\i)
            -- (dotb\i)
            -- (x\i);

        % Two halves of the reconstructed codeword
        \draw[dspconn]
            (x\i)
            -- (x\i.south |- dint\i.north);

        \draw[dspconn]
            (dotb\i)
            -- (dotb\i.south |- dint\i.north);

        % Candidate to minimum-distance selection
        \draw[dspconn]
            (dint\i)
            -- (dint\i.south |- d.north);
    }

    % Final decoded codeword
    \draw[dspconn] (d) -- (end);

\end{tikzpicture}
    \caption{Block diagram of the proposed \ac{AE-DISC} decoding.}
    \label{fig:SPAE_fig}
\end{figure}
Performance requirements vary across applications. For example, \ac{AR/VR} emphasizes low latency and robotic surgery demands exceptional reliability. To achieve both low latency and high reliability, decoders are often designed to operate under moderate-to-high \ac{SNR} conditions.
Existing decoders do not adequately meet the \ac{HRLLC} requirements. \Ac{LDPC} \ac{LMS} decoders~\cite{Ryan2005AnIT} deliver near-capacity performance for long block lengths, but their computational complexity and latency remain too high at short block lengths. Specialized short-block \ac{LDPC} \ac{LMS} decoders
~\cite{LDPC_14ns} exhibit variable latency, making it difficult to guarantee stringent latency requirements. Polar codes are well suited for short block-length scenarios~\cite{geiselhart2021polaraed}, but their low-complexity \ac{SC} decoding cannot meet the reliability and low-latency requirements of \ac{HRLLC}.
While \ac{SCL}~\cite{sarkis2014fast, sarkis2015fast, tal2015list,fan2015low,hashemi2017fast,alamdar2011simplified,balatsoukas2015llr,cocskun2022information} decoders and \ac{AE-SC}~\cite{geiselhart2021automorphism,geiselhart2021polaraed} improve reliability, they still retain the serial traversal of component decoders that limits their achievable latency. Parallel \ac{SC} architectures~\cite{li2013parallel} increase throughput by processing subcodes concurrently. However, their effectiveness depends heavily on code construction and decoding order, and improper sequencing can result in performance worse than conventional \ac{SC}. The need to synchronize parallel sub-decoders before assembling the final codeword introduces additional overhead, and the difficulty of integrating these architectures with fast-node decoding limits their practical latency~\cite{sc_parallel_architect}.
Alternative decoding methods of Polar codes are \ac{BP}-based decoding \cite{Arikan2008BP,arikan2010BP,forney2001codesongraphs}, which computes all processing elements within each stage in parallel, avoiding the sequential ordering of \ac{SC}-based approaches. However, \ac{BP} decoding typically requires a large number of iterations to achieve desirable \ac{FER} performance, and early stopping techniques introduce variable decoding latency. Existing studies have improved the latency, hardware efficiency, and error-rate performance of Polar BP decoding~\cite{bp_polar_architecture,elkelesh2018bp}, but the combination of high iteration counts and unpredictable latency still limits their applicability to \ac{HRLLC} applications.

The recently proposed \ac{Spec-SC} decoder~\cite{seah2026speculative} combines ideas from \ac{SC} and \ac{BP} decoding by introducing a new level of parallelism where both branches of the decoding tree are processed concurrently. Because the right branch operates directly on less reliable parent-node \acp{LLR}, its result is verified using an acceptance condition that triggers recomputation when the result is deemed insufficient. Although speculation reduces average latency and improves throughput, e.g., for data channel applications, the resulting variable latency is again undesirable for \ac{HRLLC}.

To address this gap, we propose the \ac{AE-DISC} decoder, which adapts the speculative parallelism of \ac{Spec-SC} to the \ac{AED} framework. Unlike \ac{Spec-SC}, the \ac{DISC} constituent decoder performs concurrent execution only at the root node and does not require recomputation, resulting in a deterministic decoding latency of approximately half that of conventional \ac{FSSC} decoding. To preserve the error-correction performance, multiple decoding attempts are performed in parallel using codeword permutations from the code's automorphism group.

By exploiting hardware parallelism, the architecture reduces decoding latency while preserving high error-correction performance, at the cost of increased computational resources. This design is particularly well suited to low-latency, high-reliability applications, namely \ac{HRLLC}, where minimizing decoding delay and maintaining a low error rate are primary design objectives. 

The main contributions of this paper are as follows:
\begin{itemize}
\item A novel decoding algorithm, namely \ac{DISC}, which has lower latency and lower complexity than standard \ac{FSSC} decoding for Polar and \ac{RM} codes.
\item \Ac{AE-DISC} decoding algorithm, which integrates \ac{DISC} decoding into the \ac{AED} framework to achieve low decoding latency and strong error-correction performance.
\item A latency analysis of \ac{AE-DISC}, following the analytical framework in~\cite{sarkis2014fast}, to quantify the latency reduction enabled by the proposed decoding architecture. %

\item A computational-complexity analysis of \ac{AE-DISC} relative to \ac{FSSC} and \ac{AE-SC} decoding for Polar and \ac{RM} codes, as well as to \ac{LDPC} decoders employing the \ac{LMS} algorithm.

\item A hardware architecture and 65-nm CMOS implementation of the proposed \ac{AE-DISC} decoder, using 5-bit \ac{LLR} quantization and 6-bit Euclidean-distance quantization, to demonstrate its low-latency advantage.

\end{itemize}

The following notation is used throughout the paper:
\begin{itemize}
\item Bold uppercase letters (e.g., $\boldsymbol{M}$) denote matrices.
\item Bold lowercase letters (e.g., $\boldsymbol{m}$) denote vectors.%
\item $\text{sgn}(\cdot)$ denotes the sign function.
\item $| \cdot |$ denotes absolute value.
\item $\oplus$ denotes the XOR operation.
\end{itemize}

\section{Preliminaries}
In this section, we introduce the necessary preliminaries on Polar codes and \ac{RM} codes. We then explain the \ac{SC}, \ac{FSSC}, and \ac{AE-SC} decoding algorithms, followed by the computational-complexity analysis of \ac{FSSC} and \ac{LMS} decoding.

\subsection{Polar Codes}
A Polar code~\cite{polar_codes_arikan} of length $N = 2^m$ and $K$ information bits is denoted by $\mathcal{P}(N,K)$. For a given construction, the input vector $\boldsymbol{u} \in \mathbb{F}_2^N$ is partitioned into two subsets: the information bits $\boldsymbol{u}_{\mathcal{I}} \in \mathbb{F}_2^K$, which occupy the most reliable polarized bit-channels, and the frozen bits $\boldsymbol{u}_{\mathcal{F}}$, which fill the remaining $N-K$ positions and are fixed to predetermined values, typically zeros. 
The index sets $\mathcal{I}$ and $\mathcal{F}$ are called the information and frozen set, respectively.
The code rate is defined as $R = K / N$.
For Polar codes constructed for \ac{SC}-based decoders, the bit-channels follow the \ac{UPO} in terms of their reliability \cite{bardet2016algebraic}. Therefore, a Polar code is compactly specified by a small set of generators $\mathcal{I}_\mathrm{min}$ that induces the remaining information indices using the \ac{UPO} \cite{geiselhart2021automorphism}.

The encoded codeword is generated as
\begin{align}
\boldsymbol{x} = \boldsymbol{u} \boldsymbol{G}_N,
\end{align}
where $\boldsymbol{G}_N = \boldsymbol{F}^{\otimes m}$ represents the $m$-fold Kronecker power of the polarization kernel,  $\boldsymbol{F} = \left[\begin{smallmatrix} 1 & 0 \\ 1 & 1\end{smallmatrix}\right]$. 

\subsection{Reed--Muller Codes}
\label{subsec:rm_codes}
Traditionally, \ac{RM} codes are formulated as binary polynomial
evaluation codes. Let $m$ be a nonnegative integer, and let
$\mathbb{F}_2[z_0,\ldots,z_{m-1}]$ denote the set of multivariate
polynomials over $\mathbb{F}_2$. Let
$\boldsymbol{z}_0,\ldots,\boldsymbol{z}_{2^m-1}$ denote the elements of
$\mathbb{F}_2^m$ in lexicographic order. For a polynomial
$p \in \mathbb{F}_2[z_0,\ldots,z_{m-1}]$, its evaluation vector is defined as
\begin{equation}
    \operatorname{eval}(p)
    =
    \bigl(
        p(\boldsymbol{z}_0),
        \ldots,
        p(\boldsymbol{z}_{2^m-1})
    \bigr).
\end{equation}

The \ac{RM} code of order $r$, denoted by
$\operatorname{RM}(r,m)$, is the set of evaluation vectors of all
binary polynomials with total degree at most $r$:
\begin{equation}
    \operatorname{RM}(r,m)
    =
    \left\{
        \operatorname{eval}(p)
        :
        \deg(p)\leq r
    \right\}.
\end{equation}

For $0\leq r\leq m$, the code parameters are \cite{macwilliams1977theory}
\begin{equation}
    N=2^m,
    \qquad
    k(r,m)=\sum_{i=0}^{r}\binom{m}{i},
    \qquad
    d_{\min}=2^{m-r}.
\end{equation}

When $r<0$, the code contains only the all-zero codeword. Moreover,
$\operatorname{RM}(0,m)$ is a repetition code,
$\operatorname{RM}(m-1,m)$ is a \ac{SPC} code, and
$\operatorname{RM}(r,m)=\mathbb{F}_2^{2^m}$ for $r\geq m$.

\ac{RM} codes can also be interpreted as a special case of Polar codes, i.e., constructed by selecting rows from the Polar transform matrix $\boldsymbol{G}_N$.
Let $\boldsymbol{b}_k$ denote the length-$m$ binary representation of row
index $k$ and $\operatorname{wt}_{\mathrm{H}}$ denote its Hamming weight. The \ac{RM} codes information set is
\begin{equation}
    \mathcal{I}_{\operatorname{RM}(r,m)}
    =
    \left\{
        k:
        \operatorname{wt}_{\mathrm{H}}(\boldsymbol{b}_k)
        \geq m-r
    \right\}.
\end{equation}

Therefore, RM and Polar codes use the same transform matrix but apply
different row-selection rules. RM codes select rows according to the
Hamming weight of their indices, whereas Polar codes select rows based
on bit-channel reliability.

\subsection{Successive Cancellation (SC)}
\label{subsec:scdecoding}
Polar and \ac{RM} codes can be efficiently decoded using the \ac{SC} algorithm, originally proposed by Schnabl and Bossert \cite{Schnabl1995} for \ac{RM} codes, and applied to Polar codes by Ar{\i}kan~\cite{polar_codes_arikan}.
Let $l_{i,t}$ denote the \ac{LLR} associated with node $i$ at stage $t$, where $ 0 \leq t \leq m$ and $0 \leq i \leq 2^m - 1$. The stage index $t$ represents the level in the decoding graph, with $t = m$ corresponding to the channel inputs and $t = 0$ to the bit decisions. The received channel LLRs are denoted by ${l_{i,m}}$.

At each stage, two types of \ac{LLR} updates are performed: one for the \textit{left branch} and one for the \textit{right branch}. The left branch combines two \acp{LLR} from the next stage using the $f$-function
\begin{equation}
\label{eq:f-minsum}
{l_{i,t}} = f\!\left({l_{i,t+1}},\, {l_{i+2^t,t+1}}\right),
\end{equation}
where \( f(\cdot) \) is defined under the min approximation as
\begin{equation}
\label{eq:f}
f(a,b) = \operatorname{sgn}(a)\,\operatorname{sgn}(b)\,
                 \min\!\big(|a|,\,|b|\big).
\end{equation}

The right branch then updates its \ac{LLR} estimate using the previously decoded left-branch decision $u_{i,t}$ through the $g$-function
\begin{equation}
\label{eq:g}
{l_{i+2^t,t}} = g\!\left( {l_{i,t+1}},\, {l_{i+2^t,t+1}, u_{i,t}}\right),
\end{equation}
where
\begin{equation}
\label{eq:gdef}
g(a,b,u) = (1 - 2u)a + b.
\end{equation}
\Cref{eq:f} and \Cref{eq:gdef} are defined for scalar inputs and extended element-wise when applied to vectors.

At leaf stage ($t=0$) the decision is made according to
\begin{equation}
u_{i,0} =
\begin{cases}
0, & \text{if } i \in \mathcal{F}; \\
\operatorname{hard}(l_{i,0}), & \text{if } i \in \mathcal{I},
\end{cases}
\end{equation}
where $\operatorname{hard}(x)$ is $1$ for $x<0$ and $0$ otherwise.
The hard decisions are propagated upwards with
\begin{equation}
u_{i,t} =  \begin{cases}
u_{i,t-1} \oplus u_{i+2^t,t-1}, & \text{if } i < 2^t, \\[6pt]
u_{i+2^t,t-1}, & \text{otherwise.}
 \end{cases} %
\label{eq:u_update}
\end{equation}

This recursive process continues until all $N = 2^m$ bits are decoded. 
In the \ac{RM} decoding literature, \ac{SC} decoding is also called \ac{GMC} decoding, however, for consistency, we refer to it as \ac{SC} decoding.
Note that the min-sum formulation of \ac{SC} decoding is invariant to scaling the input by any positive factor. Assuming \ac{BPSK} modulation $\boldsymbol{s} = \boldsymbol{1} - 2\boldsymbol{x}$ and a Gaussian channel, the channel \acp{LLR} $l_{i,m} \propto y_i$ and thus, $\boldsymbol{y}$ can be directly passed to the \ac{SC} decoder without scaling.

\subsection{Fast Simplified Successive Cancellation (FSSC) Decoding}
To improve decoding throughput, the \ac{SC} tree traversal is pruned, resulting in \ac{FSSC}~\cite{simplfiedSC, sarkis2014fast, fast_scl}. Here, special, easy-to-decode nodes in the decoding tree are classified based on their information bit patterns, where 1 denotes an information bit and 0 denotes a frozen bit. In this paper, we consider the following fast node types:
\begin{enumerate}
    \item \makebox[4.1cm][l]{\textbf{\Acf{R1}}}  : $(1, 1, \dots, 1, 1)$
    \item \makebox[4.1cm][l]{\textbf{\Acf{R0}}}  : $(0, 0, \dots, 0, 0)$
    \item \makebox[4.1cm][l]{\textbf{\Acf{SPC}}} : $(0, 1, \dots, 1, 1)$
    \item \makebox[4.1cm][l]{\textbf{\Acf{REP}}} : $(0, 0, \dots, 0, 1)$
    \item \makebox[4.1cm][l]{\textbf{First-order RM}} : $\mathrm{RM}(1,m)$
\end{enumerate}

These fast nodes can be decoded directly on the incoming \acp{LLR} without traversing the entire \ac{SC} decoding tree, thereby reducing complexity, latency, and improving throughput compared to conventional \ac{SC} decoding.
For example, \ac{SPC} constituent codes can be decoded directly using the Wagner algorithm, and first-order \ac{RM} codes are decoded using the ``Green Machine'' based on the \ac{FHT} \cite{macwilliams1977theory}.
We denote the set of fast node decoders for Polar codes as $\mathcal{T}_\mathrm{polar}=\{\mathrm{R0},\mathrm{R1},\mathrm{REP},\mathrm{SPC}\}$ and for \ac{RM} codes as $\mathcal{T}_\mathrm{RM}=\{\mathrm{SPC},\mathrm{RM1}\}$.

\subsection{Automorphism Ensemble Decoding}
\label{subsec:aed}

The automorphism group of Polar codes contains the \ac{BLTA} group defined by the equation
\begin{equation}
\boldsymbol{z}' = \boldsymbol{A}\boldsymbol{z} + \boldsymbol{b},
\label{eq:affine}
\end{equation}
where \(\boldsymbol{A}\) is an invertible, binary block-lower triangular matrix, and \(\boldsymbol{b}\) is an arbitrary binary vector. Furthermore, \(\boldsymbol{z}\) and \(\boldsymbol{z}'\) denote the binary representations of the bit indices before and after permutation, respectively. 
The block profile of \(\boldsymbol{A}\) is dependent on the code design and corresponds to the index permutations that keep the information set $\mathcal{I}$ unchanged \cite{geiselhart2021polaraed}.

For an \(\mathrm{RM}(r,m)\) code, the corresponding automorphisms are given by the general affine group, denoted by \(\mathrm{GA}(m,2)\) \cite{macwilliams1977theory}. This is a special case of \eqref{eq:affine}, where the block profile only contains a single block, and, thus, \(\boldsymbol{A}\) can be any invertible binary \(m \times m\) matrix.

Let $\pi_1, \ldots, \pi_P$ be $P$ different permutation automorphisms of the code $\mathcal{C}$ and 
$\boldsymbol{y} \in \mathbb{R}^N$ be the received vector, assuming \ac{BPSK} modulation.
A set of $P$ decoders can each decode $\pi_j(\boldsymbol{y})$ independently, before de-permuting the result with $\pi_j^{-1}$~\cite{geiselhart2021automorphism}. 
The candidate codeword decoded using $\pi_j$ is
\begin{equation}
\boldsymbol{x}_j = \pi_j^{-1}\!\big(\mathrm{Dec}(\pi_j(\boldsymbol{y}))\big).
\label{inv_perm}
\end{equation}

For Gaussian channels, the final decoding result is selected from the list of candidates according to the minimum Euclidean distance criterion
\begin{equation}
\hat{\boldsymbol{x}} = \argmin_{j = 1, \ldots, P} \| (\boldsymbol{1}-2\boldsymbol{x}_j) - \boldsymbol{y} \|.
\label{eq:aed_choice}
\end{equation}
If \ac{SC} decoders are used as constituent decoders, this variant of \ac{AED} is referred to as \ac{AE-SC}. Due to decoder equivariance, \ac{AED} of Polar codes typically requires \emph{symmetric} Polar code design with a suitable block profile~\cite{geiselhart2021polaraed}, whereas \ac{RM} codes are naturally well-suited for \ac{AED} \cite{geiselhart2021automorphism}.

\subsection{Complexity Analysis}
\label{ssec:prelim_complexity}

Following the complexity-analysis framework in~\cite{3gppR1_2507061_huawei}, the computational complexity of \ac{FSSC} decoding for Polar and \ac{RM} codes, as well as \ac{LDPC} decoding using the \ac{LMS} algorithm, can be computed by counting \ac{LLR} arithmetic and comparison operations, assuming comparable unit cost.

\subsubsection{Polar Codes and \ac{RM} Codes}

A conventional \ac{SC} decoder requires
\begin{equation}
N_{\text{op,SC}} = N\log_2 N
\end{equation}
\ac{LLR} operations. \Ac{FSSC} reduces this complexity by terminating the \ac{SC} recursion at constituent nodes that admit simplified decoding rules.
For a constituent node of length \(2^t\), conventional \ac{SC} decoding requires
\begin{equation}
C_{\text{SC}}^{(t)} = t2^t
\end{equation}
operations. The resulting \ac{FSSC} complexity can therefore be expressed as
\begin{align}
N_{\text{op,FSSC}}
&= N\log_{2}N
-\sum_{T\in\mathcal{T}}
\sum_{t=1}^{\log_2 N}
N_T^{(t)}
\left(C_{\text{SC}}^{(t)}-C_T^{(t)}\right),
\label{eq:Nop_FSSC}
\end{align}
where \(N_T^{(t)}\) denotes the number of terminal type-\(T\) constituent nodes of length \(2^t\) in the pruned decoding tree, and \(C_T^{(t)}\) denotes their direct-decoding cost.

For Polar codes, \ac{R0} and \ac{R1} nodes require no \ac{LLR} arithmetic, while a length-\(2^t\) \ac{REP} node requires \(2^t-1\) additions and an \ac{SPC} node requires \(2^t-1\) minimum operations to identify the least-reliable bit~\cite{3gppR1_2507061_huawei, krieg2025long}. For \ac{RM} codes, a first-order \(\operatorname{RM}(1,t)\) constituent is decoded using the Green Machine algorithm. Its length-\(2^t\) fast Hadamard transform requires \(t2^t\) additions or subtractions, followed by \(2^t-1\) comparisons to identify the maximum-magnitude output. Hence,
\begin{align}
C_{\mathrm{RM1}}^{(t)}
&= t2^t + (2^t-1)
= (t+1)2^t-1.
\label{eq:gmc_fht_complexity}
\end{align}

\subsubsection{\ac{LDPC} \ac{LMS} Decoding}
The computational complexity of the \ac{LDPC} \ac{LMS} decoder can be estimated by the total number of operations performed across all decoding layers and iterations. 
It is expressed as
\begin{equation}
N_{\text{op,LDPC}} = I \cdot M \cdot (4d_{c} - 1),
\label{eq:ldpc_operations}
\end{equation}
where \(I\) is the number of decoding iterations, \(d_c\) is the average check node degree, and \(M\) is the number of check nodes (after lifting)~\cite{3gppR1_2507061_huawei}.

\section{\ac{AE-DISC} Decoding}

\subsection{Algorithm Description}

\begin{algorithm}[tb]
\caption{AE-DISC Decoding}
\label{alg:spae}
\KwIn{$\boldsymbol{y}$}
\KwOut{$\hat{\boldsymbol{x}}$}
\For{$j \gets 1$ \KwTo $P$}{
    $(\boldsymbol{y}^{(L)},\, \boldsymbol{y}^{(R)}) \gets \pi_j(\boldsymbol{y})$\;

    \hspace*{-0.9em}
    $\left.
    \begin{array}{l}
      \boldsymbol{x}^{(L)} \gets \text{FSSC}(f(\boldsymbol{y}^{(L)},\boldsymbol{y}^{(R)})) \\[3pt]
      \boldsymbol{x}^{(R)} \gets \text{FSSC}(\boldsymbol{y}^{(R)})
    \end{array}
    \right\}
    \vcenter{
      \hbox{\textit{Parallel execution}}
    }$\;\vspace{-.5em}
    $\boldsymbol{x}_j \gets \pi_j^{-1}\!\left( (\boldsymbol{x}^{(L)} \oplus \boldsymbol{x}^{(R)},\,\boldsymbol{x}^{(R)}) \right)$\;
}
$\hat{\boldsymbol{x}} = \argmin\limits_{j = 1, \ldots, P} \| (\boldsymbol{1}-2\boldsymbol{x}_j) - \boldsymbol{y} \|$\;
\end{algorithm}

Previous works have proposed different scheduling inside the \ac{SC} decoding to enable new types of ensemble decoders for Polar and \ac{RM} codes \cite{bossert2025hiddencodewords,kamenev2023highratefirst}.
However, these algorithms still exhibit high latency due to the dependency of the branches.
To overcome the inherently sequential nature of \ac{SC} decoding, where each decision depends on previously decoded bits, we propose the \ac{DISC} decoder, inspired by the speculative parallelism introduced in Spec-SC~\cite{seah2026speculative}.
The component decoder parallelizes the decoding of the first and the second branch in the first stage of \ac{SC} decoding, i.e., for $t=m$. For the second branch, only the second half of the channel \acp{LLR} are used, i.e., \eqref{eq:g} becomes
\begin{equation}
\label{eq:gspea}
{l_{i+2^{m-1},m-1}} = l_{i+2^{m-1},m}.
\end{equation}
Note that this is possible as the construction of Polar codes ensures that the second half of the codeword is a valid codeword of the right component code.
Effectively, the update of the first stage is equivalent to that of \ac{BP} decoding \cite{forney2001codesongraphs, arikan2010BP} without any prior knowledge from the component codes.
As the decoding of the second branch is started immediately, we call this component algorithm \emph{\acf{DISC}} decoding.
However, because the right branch is decoded from \(\boldsymbol{y}^{(R)}\) alone, its input \acp{LLR} omit the information normally incorporated by the \(g\)-update; consequently, the error-correction performance of a standalone DISC decoder is expected to degrade. Hence, to compensate for error-correction performance loss, we utilize the \ac{AED} framework, which decodes the same received sequence using multiple code permutations, followed by selection of the codeword with the minimum Euclidean distance.
Due to the parallel nature of ensemble decoding, \ac{AE-DISC} decoding preserves the latency advantage of its constituent decoder and can theoretically reduce latency by up to half compared to conventional \ac{SC} decoding and \ac{AE-SC}.
The algorithm is summarized in~\Cref{alg:spae} and illustrated as a block diagram in \cref{fig:SPAE_fig}, where each of the $P$ branches applies a permutation and a \ac{DISC} decoder, which decodes the left and right 
constituent codes concurrently with the \ac{FSSC} algorithm. The inverse permutation is then applied before the final selection. 

\textbf{Remark:} The proposed parallelization is not restricted to the root node, and applying it at nodes deeper in the decoding tree can reduce latency further.
However, the split at the root node yields the largest latency reduction of any single split, whereas each additional split may require a larger ensemble to reach the target error-rate performance.
In this paper, we therefore restrict the parallelization to the root node.

\subsection{Path Metric Computation for AE-DISC}
To reduce computational complexity, it can be beneficial to use a \ac{PM} instead of the Euclidean distance for candidate selection \cite{kestel2023automorphism}.
In conventional \ac{SC} decoding, the \ac{PM} can be recursively computed from the signs of the \acp{LLR} at the frozen bits \cite{balatsoukas2015llr} as
\begin{align}
    \mathrm{PM}_\mathrm{SC} &= \sum_{i=0}^{N-1} (x_i \ne \operatorname{hard}(l_{i,m}) )|l_{i,m}|\label{eq:pm_full}\\
    &= \sum_{i\in \mathcal{F}} (\operatorname{hard}(l_{i,0})\ne0)|l_{i,0}|.
\end{align}
However for \ac{AE-DISC}, the right subtree is decoded from $\boldsymbol{y}^{(R)}$ alone, so the original \ac{PM} no longer holds.
Let $\mathrm{PM}^{(R)}$ denote the \ac{PM} from decoding the right half $\boldsymbol{y}^{(R)}$. From \cite{balatsoukas2015llr}, we know
\begin{align}
    \mathrm{PM}^{(R)} = \sum_{i=\nicefrac{N}{2}}^{N-1} (x_i \ne \operatorname{hard}(l_{i,m}) )|l_{i,m}|,
\end{align}
which is the second half of \eqref{eq:pm_full}. We therefore discard the left-subtree \ac{PM}, and score the left half of the candidate
directly at the root:
\begin{equation}
\begin{aligned}
\mathrm{PM}_\mathrm{DISC}=\widetilde{\mathrm{PM}}^{(L)}+\mathrm{PM}^{(R)},
\end{aligned}
\label{eq:pm-disc}
\end{equation}
where
\begin{equation}
  \widetilde{\mathrm{PM}}^{(L)} = \sum_{i=0}^{\nicefrac{N}/{2}-1}
    \left(x^{(L)}_i\oplus x^{(R)}_i\oplus \operatorname{hard}(l_{i,m})\right)\,
    |l_{i,m}|.
  \label{eq:delta_root}
\end{equation}
Instead of using the Euclidean distance, the decoder selects
\begin{equation}
  \hat{\boldsymbol{x}} = \boldsymbol{x}_{j^\star}, \qquad
  j^\star = \arg\min_{j=1,\dots,P} \mathrm{PM}_j .
  \label{eq:pm-select}
\end{equation}
Compared to Euclidean distance, only half of the bits scored at the end, while $\mathrm{PM}^{(R)}$ is accumulated inside the right decoder. This reduces the width of the adders in the selection unit and thereby shortens the critical path. 

\subsection{Computational Complexity}
\label{ssec:aedisc_complexity}
The complexity analysis of the proposed \ac{AE-DISC} decoder for Polar codes can be adapted from~\cite{3gppR1_2507061_huawei} and \Cref{eq:Nop_FSSC} as

\begin{equation}
\begin{aligned}
N_{\text{op, AE-DISC}}
&= P\cdot\left(N_{\text{op, DISC}} + N\right) \\
&= P\cdot\left(N_{\text{op, FSSC}} + \frac{N}{2}\right).
\end{aligned}
\end{equation}

The additional $+N$ terms compared to \ac{DISC} operation counts stem from the calculating the Euclidean distance metrics and comparing them among \(P\) candidate codewords.

\subsection{Decoding Latency}
\label{ssec:latency}
Latency reduction is one key advantage of \ac{AE-DISC} decoder and this section is aimed to estimate the latency reduction. To facilitate the analysis of the latency given a code construction, the following equation adapted from \cite{3gppR1_2507061_huawei, tong2023fast_latency} will be used. All \ac{R0} nodes can be completely pruned and contribute no latency.
The decoding latency \(\Phi\) depends on the number of decodable fast nodes, i.e.,
\begin{equation}
\begin{aligned}
\Phi_{s}^{(\mathrm{Polar})}(t)
&= N_{\mathrm{R1},s}(t)
 + N_{\mathrm{SPC},s}(t)
 + N_{\mathrm{REP},s}(t),\\
\Phi_{s}^{(\mathrm{RM})}(t)
&= N_{\mathrm{SPC},s}(t)
 + N_{\mathrm{RM1},s}(t).
\end{aligned}
\label{eq:fast_nodes_number}
\end{equation}
where all node counts are evaluated at stage~$t$, and
$s\in\{\mathrm{L},\mathrm{R}\}$ denotes the decoding side.

The total decoding latency $T_{\text{L}}$ of the left and $T_{\text{R}}$ of the right decoding tree the can be decomposed into the tree-traversing latency and the latency of the fast nodes as
\begin{equation}
T_{\text{L}}=T_{\text{tree}}^{\mathrm{L}}\!\bigl(N,\mathrm{PE}_{\mathrm{L}}\bigr)
+ \sum_{t=1}^{\log_{2} N} \!\Phi_{\mathrm{L}}(t),
\label{eq:T_left}
\end{equation}
\begin{equation}
T_{\text{R}}=T_{\text{tree}}^{\mathrm{R}}\!\bigl(N,\mathrm{PE}_{\mathrm{R}}\bigr)
+ \sum_{t=1}^{\log_{2} N} \!\Phi_{\mathrm{R}}(t),
\label{eq:T_right}
\end{equation}
where \({T_{\text{tree}}^s}(N, \text{PE}_{\text{s}})\) is the decoding time in the binary decoding tree excluding the processing time at the decision leaves, given the hardware constraint of the number  of processing elements \(\text{PE}_{\text{s}}\). 
The overall latency of the \ac{AE-DISC} decoding can be computed as
\begin{equation}
\label{eq:T_spae}
\begin{split}
T_{\text{AE-DISC}}
&= \max\!\big\{
T_{\text{L}},\,
T_{\text{R}}
\big\}
+ \delta,
\end{split}
\end{equation}
where $\delta$ denotes the constant overhead associated with the final combine operation at the root node and the selection of the best codeword, where additional alignment or switching latency may occur between parallel decoding paths.

For the \ac{AE-DISC} decoder, the total latency is bounded by the slower of left and right branch rather than their sum, leading to a theoretical speed-up bounded by
\begin{equation}
\label{eq:bound}
1 < 
\text{speed-up}
= \frac{T_{\text{AE-SC}}}{T_{\text{AE-DISC}}}
\leq 2.
\end{equation}

Speed-up is defined as the ratio of the baseline decoding time, in this case \ac{AE-SC}, to the decoding time of the proposed \ac{AE-DISC}, and thus directly reflects the relative latency reduction. The speed-up approaches $2\times$ when both halves exhibit comparable decoding cycles. In this case, a speed-up of \(2\times\) roughly corresponds to halving the average decoding latency, assuming similar clock frequency and hardware utilization efficiency. 

For simplicity, we assume each fast-decodable node of type can be processed in one time unit. R0 type fast node does not require decoding time. Also we assume we are not limited by processing element number, i.e., a fully parallel decoding architecture. The maximum number of bits to be decided in parallel is equal to \(N\). The tree-traversal time excluding fast nodes can be estimated by counting the number of operations including \(g, f\) and \(c\).
 
\Cref{tab:speedup} lists the computed speed-up for various Polar and \ac{RM} codes of different rates and lengths. 
For the \( \mathcal{P}(128,64) \) code, a speed-up of \(1.93\times\) is achieved, while for the \( \mathcal{P}(256,128) \) code, a speed-up of \(1.95\times\) is obtained. For the high-rate \( \mathcal{P}(128,98) \) and low-rate \( \mathcal{P}(128,34) \) codes, the speed-up is lower, at \(1.46\times\) and \(1.45\times\), respectively. This is because the \ac{AE-DISC} latency is bounded by the slower of the two parallel halves, and at these rates the left and right operations are highly unbalanced. 
For the \ac{RM} codes, \ac{AE-DISC} achieves speed-ups of \(1.61\times\)
for \(\mathrm{RM}(3,7)\) and \(1.93\times\) for \(\mathrm{RM}(4,8)\). 
Note that the speed-up is independent of the
ensemble size because all component \ac{DISC} decoders are operating in parallel,
assuming the candidate-selection time overhead is small.

\begin{table*}[t]
\centering
\def\arraystretch{1.35}
\caption{Speed-up of \ac{AE-DISC} compared with \ac{AE-SC}.}
\begin{tabular}{cccccccccc}
\hline
Code & $\mathcal{I}_{\mathrm{min}}$ & $N$ & $K$ & $R$
& Left Op & Right Op & AE-SC & AE-DISC & Speed-up \\
\hline

Polar & $\{23,112\}$ & 128 & 64 & 0.5
& 37 & 38 & 77 & 40 & 1.93 \\

Polar & $\{11\}$ & 128 & 98 & 0.77
& 37 & 18 & 57 & 39 & 1.46 \\

Polar & $\{31,60\}$ & 128 & 34 & 0.27
& 18 & 38 & 58 & 40 & 1.45 \\

Polar & $\{31,57\}$ & 256 & 128 & 0.5
& 38 & 38 & 78 & 40 & 1.95 \\

$\mathrm{RM}(3,7)$ & $\{15\}$ & 128 & 64 & 0.5
& 14 & 21 & 37 & 23 & 1.61 \\

$\mathrm{RM}(4,8)$ & $\{15\}$ & 256 & 163 & 0.64
& 38 & 37 & 77 & 40 & 1.93 \\

\hline
\end{tabular}
\label{tab:speedup}
\end{table*}

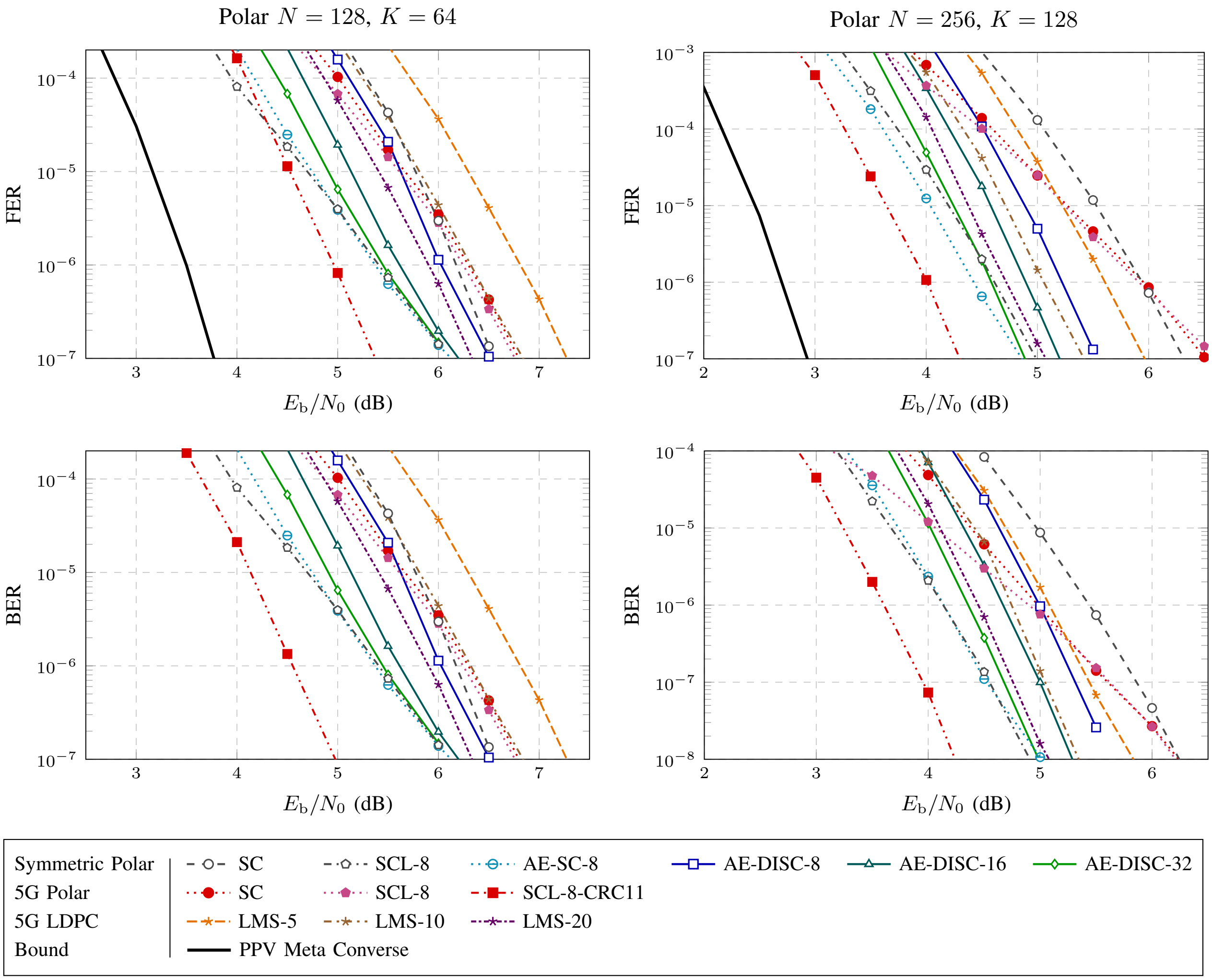
\begin{figure*}[t]
    \centering
    \subfigure{%
    \label{fig:fer_128}
    \begin{tikzpicture}

\begin{axis}[
    xlabel={$E_\mathrm{b}/N_0$ (dB)},
    ylabel={FER},
    title={Polar $N=128$, $K=64$},
    width=0.5\linewidth,
    height=0.34\linewidth,
    ymin=1e-7,
    ymax=2e-4,
    xmin=2.5,
    xmax=7.5,
    ymode=log,
    clip=true,
    legend style={
        anchor=center,
        cells={anchor=west},
        column sep=2mm,
        font=\fontsize{7pt}{7.2pt}\selectfont,
    },
    legend columns=3,
    legend to name=legend:polar_fer_128_64,
]

% --------------------------------------------------
% 5G FSSC
% --------------------------------------------------
\addplot[fssc]
table[
    x=SNR_5G,
    y=FER_5G,
    col sep=comma
]{data/aed_spae_128_64.csv};
\addlegendentry{FSSC}
\label{leg:5g-sc}

% --------------------------------------------------
% AE-DISC-8
% --------------------------------------------------
\addplot[spae8]
table[
    x=SNR_P8,
    y=FER_P8,
    col sep=comma
]{data/aed_spae_128_64.csv};
\addlegendentry{AE-DISC-8}
\label{leg:aedisc-8}

% --------------------------------------------------
% AE-DISC-16
% --------------------------------------------------
\addplot[spae16]
table[
    x=SNR_P16,
    y=FER_P16,
    col sep=comma
]{data/aed_spae_128_64.csv};
\addlegendentry{AE-DISC-16}
\label{leg:aedisc-16}

% --------------------------------------------------
% AE-DISC-32
% --------------------------------------------------
\addplot[spae32]
table[
    x=SNR_P32,
    y=FER_P32,
    col sep=comma
]{data/aed_spae_128_64.csv};
\addlegendentry{AE-DISC-32}
\label{leg:aedisc-32}

% --------------------------------------------------
% 5G SCL-8
% --------------------------------------------------
\addplot[scl8]
table[
    x=SNR_SCL8,
    y=FER_SCL8,
    col sep=comma
]{data/aed_spae_128_64.csv};
\addlegendentry{FSSCL-8}
\label{leg:5g-scl}

% --------------------------------------------------
% LDPC-5
% --------------------------------------------------
\addplot[ldpc5]
table[
    x=SNR_LDPC_5,
    y=FER_LDPC_5,
    col sep=comma
]{data/aed_spae_128_64.csv};
\addlegendentry{LDPC-5}
\label{leg:ldpc-5}

% --------------------------------------------------
% LDPC-10
% --------------------------------------------------
\addplot[ldpc10]
table[
    x=SNR_LDPC_10,
    y=FER_LDPC_10,
    col sep=comma
]{data/aed_spae_128_64.csv};
\addlegendentry{LDPC-10}
\label{leg:ldpc-10}

% --------------------------------------------------
% LDPC-20
% --------------------------------------------------
\addplot[ldpc20]
table[
    x=SNR_LDPC_20,
    y=FER_LDPC_20,
    col sep=comma
]{data/aed_spae_128_64.csv};
\addlegendentry{LDPC-20}
\label{leg:ldpc-20}

% --------------------------------------------------
% AE-SC-8
% --------------------------------------------------
\addplot[aed8]
table[
    x=SNR_AED8,
    y=FER_AED8,
    col sep=comma
]{data/aed_spae_128_64.csv};
\addlegendentry{AE-SC-8}
\label{leg:ae-sc}

% --------------------------------------------------
% Construction [23,112] + SC
% --------------------------------------------------
\addplot[construction-sc]
table[
    x=SNR_SC_23_112,
    y=FER_SC_23_112,
    col sep=comma
]{data/aed_spae_128_64.csv};
\addlegendentry{SC}
\label{leg:sym-sc}

% --------------------------------------------------
% Construction [23,112] + SCL-8
% --------------------------------------------------
\addplot[construction-scl8]
table[
    x=SNR_SCL8_23_112,
    y=FER_SCL8_23_112,
    col sep=comma
]{data/aed_spae_128_64.csv};
\addlegendentry{SCL-8}
\label{leg:sym-scl}

% --------------------------------------------------
% PPV meta converse bound
% --------------------------------------------------
\addplot[ppv]
table[
    x=SNR_PPV,
    y=FER_PPV,
    col sep=comma
]{data/aed_spae_128_64.csv};
\addlegendentry{PPV bound}
\label{leg:ppv}

%--------------------
%SCL-8 CRC11 5G
%-------
\addplot[5g-scl8-crc]
table[
x=EbN0_dB,
y=FER_128_64,
col sep=comma
]{data/SCL-8-CRC11/data.csv};
\addlegendentry{SCL-8 CRC}
\label{leg:scl8_crc}

\end{axis}

\end{tikzpicture}
    }  
    \subfigure{%
        \label{fig:fer_256}
        \begin{tikzpicture}

\begin{axis}[
    xlabel={$E_\mathrm{b}/N_0$ (dB)},
    ylabel={FER},
    title={Polar $N=256$, $K=128$},
    width=0.5\linewidth,
    height=0.34\linewidth,
    ymin=1e-7,
    ymax=1e-3,
    xmin=2,
    xmax=6.5,
    clip=true,
    scale=1,
    ymode=log,
    legend style={
        anchor=center,
        cells={anchor=west},
        column sep=2mm,
        thick,
        font=\fontsize{7pt}{7.2pt}\selectfont,
    },
    legend to name=legend:polar_fer_256_128,
    legend columns=3,
    legend pos=north east,
]

% --------------------------------------------------
% 5G FSSC
% --------------------------------------------------
\addplot[fssc]
table[
    x=SNR_5G,
    y=FER_5G,
    col sep=comma
]{data/aed_spae_256_128.csv};
\label{leg:5g-sc-256}

% --------------------------------------------------
% AE-DISC-8
% --------------------------------------------------
\addplot[spae8]
table[
    x=SNR_P8,
    y=FER_P8,
    col sep=comma
]{data/aed_spae_256_128.csv};
\label{leg:aedisc-8-256}

% --------------------------------------------------
% AE-DISC-16
% --------------------------------------------------
\addplot[spae16]
table[
    x=SNR_P16,
    y=FER_P16,
    col sep=comma
]{data/aed_spae_256_128.csv};
\label{leg:aedisc-16-256}

% --------------------------------------------------
% AE-DISC-32
% --------------------------------------------------
\addplot[spae32]
table[
    x=SNR_P32,
    y=FER_P32,
    col sep=comma
]{data/aed_spae_256_128.csv};
\label{leg:aedisc-32-256}

% --------------------------------------------------
% 5G SCL-8
% --------------------------------------------------
\addplot[scl8]
table[
    x=SNR_SCL8,
    y=FER_SCL8,
    col sep=comma
]{data/aed_spae_256_128.csv};
\label{leg:5g-scl-256}

% --------------------------------------------------
% 5G SCL-8 CRC
% --------------------------------------------------
\addplot[5g-scl8-crc]
table[
    x=EbN0_dB,
    y=FER_256_128,
    col sep=comma
]{data/SCL-8-CRC11/data.csv};
\addlegendentry{SCL-8 CRC}
\label{leg:scl8_crc}

% --------------------------------------------------
% LDPC-5
% --------------------------------------------------
\addplot[ldpc5]
table[
    x=SNR_LDPC_5,
    y=FER_LDPC_5,
    col sep=comma
]{data/aed_spae_256_128.csv};
\label{leg:ldpc-5-256}

% --------------------------------------------------
% LDPC-10
% --------------------------------------------------
\addplot[ldpc10]
table[
    x=SNR_LDPC_10,
    y=FER_LDPC_10,
    col sep=comma
]{data/aed_spae_256_128.csv};
\label{leg:ldpc-10-256}

% --------------------------------------------------
% LDPC-20
% --------------------------------------------------
\addplot[ldpc20]
table[
    x=SNR_LDPC_20,
    y=FER_LDPC_20,
    col sep=comma
]{data/aed_spae_256_128.csv};
\label{leg:ldpc-20-256}

% --------------------------------------------------
% AE-SC-8
% --------------------------------------------------
\addplot[aed8]
table[
    x=SNR_AED8,
    y=FER_AED8,
    col sep=comma
]{data/aed_spae_256_128.csv};
\label{leg:ae-sc-256}

% --------------------------------------------------
% [31,57] construction + SC
% --------------------------------------------------
\addplot[construction-sc]
table[
    x=SNR_31_57_SC,
    y=FER_31_57_SC,
    col sep=comma
]{data/aed_spae_256_128.csv};
\label{leg:sym-sc-256}

% --------------------------------------------------
% [31,57] construction + SCL-8
% --------------------------------------------------
\addplot[construction-scl8]
table[
    x=SNR_SCL8_23_112,
    y=FER_SCL8_23_112,
    col sep=comma
]{data/aed_spae_256_128.csv};
\label{leg:sym-scl-256}

% --------------------------------------------------
% PPV meta converse bound
% --------------------------------------------------
\addplot[ppv]
table[
    x=SNR_PPV,
    y=FER_PPV,
    col sep=comma
]{data/aed_spae_256_128.csv};
\addlegendentry{PPV bound}
\label{leg:ppv}

\end{axis}

\end{tikzpicture}
    }
    \subfigure{%
        \label{fig:ber_128}
        \begin{tikzpicture}

\begin{axis}[
    xlabel={$E_\mathrm{b}/N_0$ (dB)},
    ylabel={BER},
    width=0.5\linewidth,
    height=0.34\linewidth,
    ymin=1e-7,
    ymax=2e-4,
    xmin=2.5,
    xmax=7.5,
    clip=true,
    scale=1,
    ymode=log,
    legend style={
        anchor=center,
        cells={anchor=west},
        column sep=2mm,
        thick,
        font=\fontsize{7pt}{7.2pt}\selectfont,
    },
    legend to name=legend:polar_fer_128_64,
    legend columns=3,
    legend pos=north east,
]

% --------------------------------------------------
% 5G FSSC
% --------------------------------------------------
\addplot[fssc]
table[
    x=SNR_5G,
    y=FER_5G,
    col sep=comma
]{data/aed_spae_128_64.csv};
\label{leg:5g-sc}

% --------------------------------------------------
% AE-DISC-8
% --------------------------------------------------
\addplot[spae8]
table[
    x=SNR_P8,
    y=FER_P8,
    col sep=comma
]{data/aed_spae_128_64.csv};
\label{leg:aedisc-8}

% --------------------------------------------------
% AE-DISC-16
% --------------------------------------------------
\addplot[spae16]
table[
    x=SNR_P16,
    y=FER_P16,
    col sep=comma
]{data/aed_spae_128_64.csv};
\label{leg:aedisc-16}

% --------------------------------------------------
% AE-DISC-32
% --------------------------------------------------
\addplot[spae32]
table[
    x=SNR_P32,
    y=FER_P32,
    col sep=comma
]{data/aed_spae_128_64.csv};
\label{leg:aedisc-32}

% --------------------------------------------------
% 5G SCL-8
% --------------------------------------------------
\addplot[scl8]
table[
    x=SNR_SCL8,
    y=FER_SCL8,
    col sep=comma
]{data/aed_spae_128_64.csv};
\label{leg:5g-scl}

% --------------------------------------------------
% LDPC-5
% --------------------------------------------------
\addplot[ldpc5]
table[
    x=SNR_LDPC_5,
    y=FER_LDPC_5,
    col sep=comma
]{data/aed_spae_128_64.csv};
\label{leg:ldpc-5}

% --------------------------------------------------
% LDPC-10
% --------------------------------------------------
\addplot[ldpc10]
table[
    x=SNR_LDPC_10,
    y=FER_LDPC_10,
    col sep=comma
]{data/aed_spae_128_64.csv};
\label{leg:ldpc-10}

% --------------------------------------------------
% LDPC-20
% --------------------------------------------------
\addplot[ldpc20]
table[
    x=SNR_LDPC_20,
    y=FER_LDPC_20,
    col sep=comma
]{data/aed_spae_128_64.csv};
\label{leg:ldpc-20}

% --------------------------------------------------
% AE-SC-8
% --------------------------------------------------
\addplot[aed8]
table[
    x=SNR_AED8,
    y=FER_AED8,
    col sep=comma
]{data/aed_spae_128_64.csv};
\label{leg:ae-sc}

% --------------------------------------------------
% [23,112] construction + SC
% --------------------------------------------------
\addplot[construction-sc]
table[
    x=SNR_SC_23_112,
    y=FER_SC_23_112,
    col sep=comma
]{data/aed_spae_128_64.csv};
\label{leg:sym-sc}

% --------------------------------------------------
% [23,112] construction + SCL-8
% --------------------------------------------------
\addplot[construction-scl8]
table[
    x=SNR_SCL8_23_112,
    y=FER_SCL8_23_112,
    col sep=comma
]{data/aed_spae_128_64.csv};
\label{leg:sym-scl}

% --------------------------------------------------
% 5G SCL-8 CRC
% --------------------------------------------------
\addplot[5g-scl8-crc]
table[
    x=EbN0_dB,
    y=BER_128_64,
    col sep=comma
]{data/SCL-8-CRC11/data.csv};
\addlegendentry{SCL-8 CRC}
\label{leg:scl8_crc}

\end{axis}

\end{tikzpicture}
    }
    \subfigure{%
        \label{fig:ber_256}
        \begin{tikzpicture}

\begin{axis}[
    xlabel={$E_\mathrm{b}/N_0$ (dB)},
    ylabel={BER},
    width=0.5\linewidth,
    height=0.34\linewidth,
    ymin=1e-8,
    ymax=1e-4,
    xmin=2,
    xmax=6.5,
    clip=true,
    scale=1,
    ymode=log,
    legend style={
        anchor=center,
        cells={anchor=west},
        column sep=2mm,
        thick,
        font=\fontsize{7pt}{7.2pt}\selectfont,
    },
    legend to name=legend:polar_ber_256_128,
    legend columns=3,
    legend pos=north east,
]

% --------------------------------------------------
% 5G FSSC
% --------------------------------------------------
\addplot[fssc]
table[
    x=SNR_5G,
    y=BER_5G,
    col sep=comma
]{data/output_plot256_BER.csv};
\label{leg:5g-sc-ber-256}

% --------------------------------------------------
% AE-DISC-8
% --------------------------------------------------
\addplot[spae8]
table[
    x=SNR_P8,
    y=BER_P8,
    col sep=comma
]{data/output_plot256_BER.csv};
\label{leg:aedisc-8-ber-256}

% --------------------------------------------------
% AE-DISC-16
% --------------------------------------------------
\addplot[spae16]
table[
    x=SNR_P16,
    y=BER_P16,
    col sep=comma
]{data/output_plot256_BER.csv};
\label{leg:aedisc-16-ber-256}

% --------------------------------------------------
% AE-DISC-32
% --------------------------------------------------
\addplot[spae32]
table[
    x=SNR_P32,
    y=BER_P32,
    col sep=comma
]{data/output_plot256_BER.csv};
\label{leg:aedisc-32-ber-256}

% --------------------------------------------------
% 5G SCL-8
% --------------------------------------------------
\addplot[scl8]
table[
    x=SNR_SCL8,
    y=BER_SCL8,
    col sep=comma
]{data/output_plot256_BER.csv};
\label{leg:5g-scl-ber-256}

% --------------------------------------------------
% 5G SCL-8 CRC
% --------------------------------------------------
\addplot[5g-scl8-crc]
table[
    x=EbN0_dB,
    y=BER_256_128,
    col sep=comma
]{data/SCL-8-CRC11/data.csv};
\addlegendentry{SCL-8 CRC}
\label{leg:scl8_crc}

% --------------------------------------------------
% LDPC-5
% --------------------------------------------------
\addplot[ldpc5]
table[
    x=SNR_LDPC_5,
    y=BER_LDPC_5,
    col sep=comma
]{data/output_plot256_BER.csv};
\label{leg:ldpc-5-ber-256}

% --------------------------------------------------
% LDPC-10
% --------------------------------------------------
\addplot[ldpc10]
table[
    x=SNR_LDPC_10,
    y=BER_LDPC_10,
    col sep=comma
]{data/output_plot256_BER.csv};
\label{leg:ldpc-10-ber-256}

% --------------------------------------------------
% LDPC-20
% --------------------------------------------------
\addplot[ldpc20]
table[
    x=SNR_LDPC_20,
    y=BER_LDPC_20,
    col sep=comma
]{data/output_plot256_BER.csv};
\label{leg:ldpc-20-ber-256}

% --------------------------------------------------
% AE-SC-8
% --------------------------------------------------
\addplot[aed8]
table[
    x=SNR_AED8,
    y=BER_AED8,
    col sep=comma
]{data/output_plot256_BER.csv};
\label{leg:ae-sc-ber-256}

% --------------------------------------------------
% [31,57] construction + SC
% --------------------------------------------------
\addplot[construction-sc]
table[
    x=SNR_31_57_SC,
    y=BER_31_57_SC,
    col sep=comma
]{data/output_plot256_BER.csv};
\label{leg:sym-sc-ber-256}

% --------------------------------------------------
% [31,57] construction + SCL-8
% --------------------------------------------------
\addplot[construction-scl8]
table[
    x=SNR_SCL8_23_112,
    y=BER_SCL8_23_112,
    col sep=comma
]{data/output_plot256_BER.csv};
\label{leg:sym-scl-ber-256}

\end{axis}

\end{tikzpicture}
    }
    \\\vspace{1pt}
    \PolarDecoderLegend
   
    \caption{BER and FER comparison of \ac{AE-DISC} decoder to other decoding algorithms for Polar codes and \ac{LDPC} codes. }
     \label{fig:polar_fer}
\end{figure*}

\section{Simulation Results}

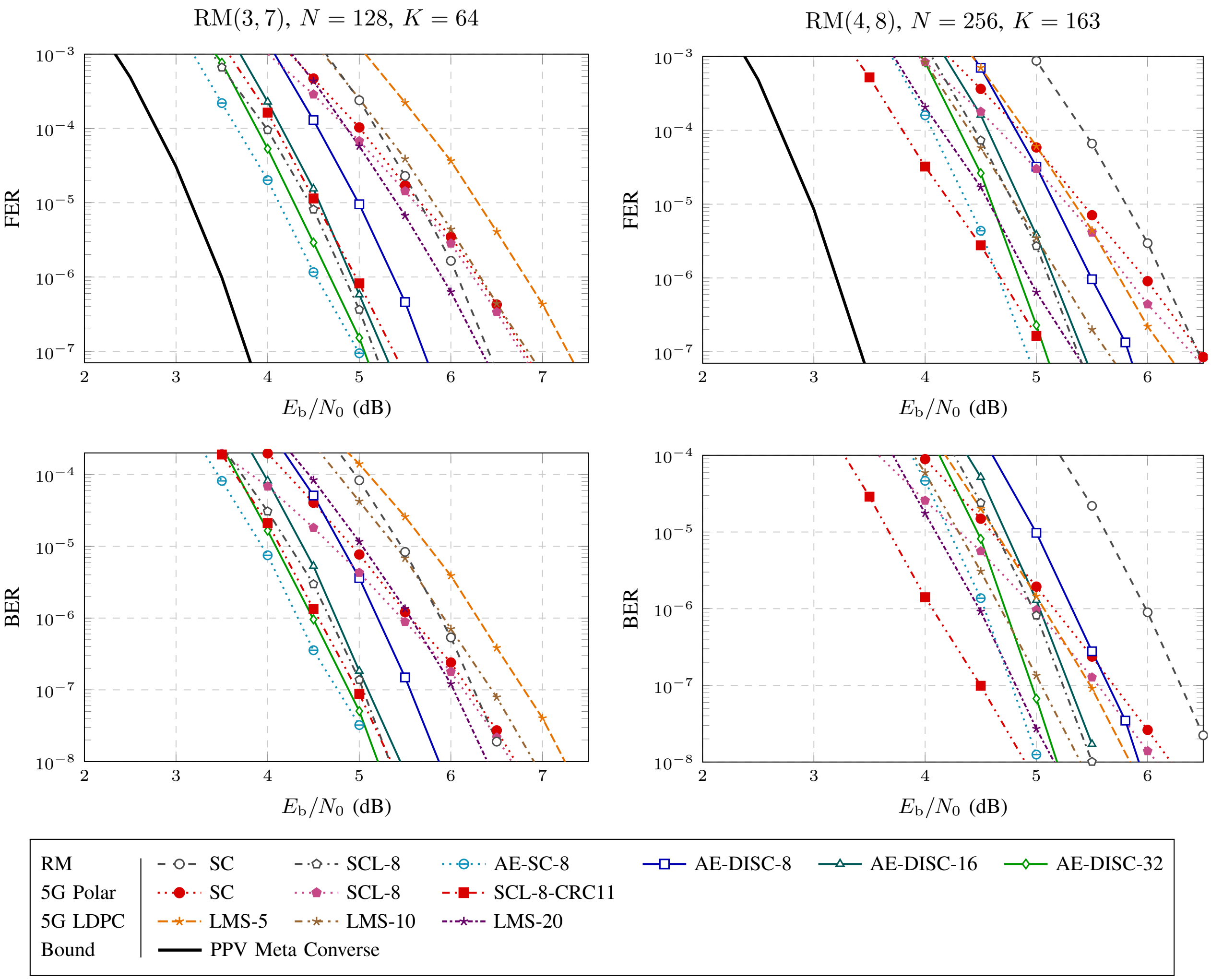
\begin{figure*}[!t]
    \centering
    \subfigure{%
        \label{fig:fer_128_rm}
        \begin{tikzpicture}

\begin{axis}[
    xlabel={$E_\mathrm{b}/N_0$ (dB)},
    ylabel={FER},
    title={$\mathrm{RM}(3,7)$, $N=128$, $K=64$},
    width=0.5\linewidth,
    height=0.34\linewidth,
    ymin=7e-8,
    ymax=1e-3,
    xmin=2,
    xmax=7.5,
    clip=true,
    scale=1,
    ymode=log,
    legend style={
        anchor=center,
        cells={anchor=west},
        column sep=2mm,
        thick,
        font=\fontsize{7pt}{7.2pt}\selectfont,
    },
    legend to name=legend:rm_fer_128_64,
    legend columns=3,
    legend pos=north east,
]

% --------------------------------------------------
% FSSC
% --------------------------------------------------
\addplot[fssc]
table[
    x=SNR_5G,
    y=FER_5G,
    col sep=comma
]{data/RM/aed_spae_128_64.csv};

% --------------------------------------------------
% AE-DISC-8
% --------------------------------------------------
\addplot[spae8]
table[
    x=SNR_P8,
    y=FER_P8,
    col sep=comma
]{data/RM/aed_spae_128_64.csv};

% --------------------------------------------------
% AE-DISC-16
% --------------------------------------------------
\addplot[spae16]
table[
    x=SNR_P16,
    y=FER_P16,
    col sep=comma
]{data/RM/aed_spae_128_64.csv};

% --------------------------------------------------
% AE-DISC-32
% --------------------------------------------------
\addplot[spae32]
table[
    x=SNR_P32,
    y=FER_P32,
    col sep=comma
]{data/RM/aed_spae_128_64.csv};

% --------------------------------------------------
% SCL-8
% --------------------------------------------------
\addplot[scl8]
table[
    x=SNR_SCL8,
    y=FER_SCL8,
    col sep=comma
]{data/RM/aed_spae_128_64.csv};

% --------------------------------------------------
% LDPC-5
% --------------------------------------------------
\addplot[ldpc5]
table[
    x=SNR_LDPC_5,
    y=FER_LDPC_5,
    col sep=comma
]{data/RM/aed_spae_128_64.csv};

% --------------------------------------------------
% LDPC-10
% --------------------------------------------------
\addplot[ldpc10]
table[
    x=SNR_LDPC_10,
    y=FER_LDPC_10,
    col sep=comma
]{data/RM/aed_spae_128_64.csv};

% --------------------------------------------------
% LDPC-20
% --------------------------------------------------
\addplot[ldpc20]
table[
    x=SNR_LDPC_20,
    y=FER_LDPC_20,
    col sep=comma
]{data/RM/aed_spae_128_64.csv};

% --------------------------------------------------
% AE-SC-8
% --------------------------------------------------
\addplot[aed8]
table[
    x=SNR_AED8,
    y=FER_AED8,
    col sep=comma
]{data/RM/aed_spae_128_64.csv};

% --------------------------------------------------
% RM construction + SC
% --------------------------------------------------
\addplot[construction-sc]
table[
    x=SNR_SC_RM,
    y=FER_SC_RM,
    col sep=comma
]{data/RM/aed_spae_128_64.csv};
\label{leg:sym-sc}

% --------------------------------------------------
% RM construction + SCL-8
% --------------------------------------------------
\addplot[construction-scl8]
table[
    x=SNR_SCL8_RM,
    y=FER_SCL8_RM,
    col sep=comma
]{data/RM/aed_spae_128_64.csv};
\label{leg:sym-scl}

%--------------------
%SCL-8 CRC11 5G
%-------
\addplot[5g-scl8-crc]
table[
x=EbN0_dB,
y=FER_128_64,
col sep=comma
]{data/SCL-8-CRC11/data.csv};
\addlegendentry{SCL-8 CRC}
\label{leg:scl8_crc}

% --------------------------------------------------
% PPV meta converse bound
% --------------------------------------------------
\addplot[ppv]
table[
    x=SNR_PPV_RM,
    y=FER_PPV_RM,
    col sep=comma
]{data/RM/aed_spae_128_64.csv};
\addlegendentry{PPV bound}
\label{leg:ppv}

\end{axis}

\end{tikzpicture}
    }
    \subfigure{%
        \label{fig:fer_256_rm}
        \begin{tikzpicture}

\begin{axis}[
    xlabel={$E_\mathrm{b}/N_0$ (dB)},
    ylabel={FER},
    title={$\mathrm{RM}(4,8)$, $N=256$, $K=163$},
    width=0.5\linewidth,
    height=0.34\linewidth,
    ymin=7e-8,
    ymax=1e-3,
    xmin=2,
    xmax=6.5,
    clip=true,
    scale=1,
    ymode=log,
    legend style={
        anchor=center,
        cells={anchor=west},
        column sep=2mm,
        thick,
        font=\fontsize{7pt}{7.2pt}\selectfont,
    },
    legend to name=legend:rm_fer_256_163,
    legend columns=3,
    legend pos=north east,
]

% --------------------------------------------------
% FSSC
% --------------------------------------------------
\addplot[fssc]
table[
    x=SNR_5G,
    y=FER_5G,
    col sep=comma
]{data/RM/aed_spae_256_163.csv};

% --------------------------------------------------
% AE-DISC-8
% --------------------------------------------------
\addplot[spae8]
table[
    x=SNR_P8,
    y=FER_P8,
    col sep=comma
]{data/RM/aed_spae_256_163.csv};

% --------------------------------------------------
% AE-DISC-16
% --------------------------------------------------
\addplot[spae16]
table[
    x=SNR_P16,
    y=FER_P16,
    col sep=comma
]{data/RM/aed_spae_256_163.csv};

% --------------------------------------------------
% AE-DISC-32
% --------------------------------------------------
\addplot[spae32]
table[
    x=SNR_P32,
    y=FER_P32,
    col sep=comma
]{data/RM/aed_spae_256_163.csv};

% --------------------------------------------------
% SCL-8
% --------------------------------------------------
\addplot[scl8]
table[
    x=SNR_SCL8,
    y=FER_SCL8,
    col sep=comma
]{data/RM/aed_spae_256_163.csv};

% --------------------------------------------------
% LDPC-5
% --------------------------------------------------
\addplot[ldpc5]
table[
    x=SNR_LDPC_5,
    y=FER_LDPC_5,
    col sep=comma
]{data/RM/aed_spae_256_163.csv};

% --------------------------------------------------
% LDPC-10
% --------------------------------------------------
\addplot[ldpc10]
table[
    x=SNR_LDPC_10,
    y=FER_LDPC_10,
    col sep=comma
]{data/RM/aed_spae_256_163.csv};

% --------------------------------------------------
% LDPC-20
% --------------------------------------------------
\addplot[ldpc20]
table[
    x=SNR_LDPC_20,
    y=FER_LDPC_20,
    col sep=comma
]{data/RM/aed_spae_256_163.csv};

% --------------------------------------------------
% AE-SC-8
% --------------------------------------------------
\addplot[aed8]
table[
    x=SNR_AED8,
    y=FER_AED8,
    col sep=comma
]{data/RM/aed_spae_256_163.csv};

% --------------------------------------------------
% RM construction + SC
% --------------------------------------------------
\addplot[construction-sc]
table[
    x=SNR_SC_RM,
    y=FER_SC_RM,
    col sep=comma
]{data/RM/aed_spae_256_163.csv};
\label{leg:sym-sc}

% --------------------------------------------------
% RM construction + SCL-8
% --------------------------------------------------
\addplot[construction-scl8]
table[
    x=SNR_SCL8_RM,
    y=FER_SCL8_RM,
    col sep=comma
]{data/RM/aed_spae_256_163.csv};
\label{leg:sym-scl}

% --------------------------------------------------
% PPV meta converse bound
% --------------------------------------------------
\addplot[ppv]
table[
    x=SNR_PPV_RM,
    y=FER_PPV_RM,
    col sep=comma
]{data/RM/aed_spae_256_163.csv};
\addlegendentry{PPV bound}
\label{leg:ppv}

%--------------------
%SCL-8 CRC11 5G
%-------
\addplot[5g-scl8-crc]
table[
x=EbN0_dB,
y=FER_256_163,
col sep=comma
]{data/SCL-8-CRC11/data.csv};
\addlegendentry{SCL-8 CRC}
\label{leg:scl8_crc}

\end{axis}

\end{tikzpicture}
    }
    \subfigure{%
        \label{fig:ber_128_rm}
        \begin{tikzpicture}

\begin{axis}[
    xlabel={$E_\mathrm{b}/N_0$ (dB)},
    ylabel={BER},
    width=0.5\linewidth,
    height=0.34\linewidth,
    ymin=1e-8,
    ymax=2e-4,
    xmin=2,
    xmax=7.5,
    clip=true,
    scale=1,
    ymode=log,
    legend style={
        anchor=center,
        cells={anchor=west},
        column sep=2mm,
        thick,
        font=\fontsize{7pt}{7.2pt}\selectfont,
    },
    legend to name=legend:rm_ber_128_64,
    legend columns=5,
    legend pos=north east,
]

% ==================================================
% Legend entries
% ==================================================

\addlegendimage{fssc}
\addlegendentry{FSSC}

\addlegendimage{spae8}
\addlegendentry{AE-DISC-8}

\addlegendimage{spae16}
\addlegendentry{AE-DISC-16}

\addlegendimage{spae32}
\addlegendentry{AE-DISC-32}

\addlegendimage{scl8}
\addlegendentry{FSSCL-8}

\addlegendimage{ldpc5}
\addlegendentry{LDPC-5}

\addlegendimage{ldpc10}
\addlegendentry{LDPC-10}

\addlegendimage{ldpc20}
\addlegendentry{LDPC-20}

\addlegendimage{aed8}
\addlegendentry{AE-SC-8}

% ==================================================
% Curves
% ==================================================

% --------------------------------------------------
% FSSC
% --------------------------------------------------
\addplot[fssc]
table[
    x=SNR_5G,
    y=BER_5G,
    col sep=comma
]{data/RM/output_plot128_BER.csv};

% --------------------------------------------------
% AE-DISC-8
% --------------------------------------------------
\addplot[spae8]
table[
    x=SNR_P8,
    y=BER_P8,
    col sep=comma
]{data/RM/output_plot128_BER.csv};

% --------------------------------------------------
% AE-DISC-16
% --------------------------------------------------
\addplot[spae16]
table[
    x=SNR_P16,
    y=BER_P16,
    col sep=comma
]{data/RM/output_plot128_BER.csv};

% --------------------------------------------------
% AE-DISC-32
% --------------------------------------------------
\addplot[spae32]
table[
    x=SNR_P32,
    y=BER_P32,
    col sep=comma
]{data/RM/output_plot128_BER.csv};

% --------------------------------------------------
% FSSCL-8
% --------------------------------------------------
\addplot[scl8]
table[
    x=SNR_SCL8,
    y=BER_SCL8,
    col sep=comma
]{data/RM/output_plot128_BER.csv};

% --------------------------------------------------
% LDPC-5
% --------------------------------------------------
\addplot[ldpc5]
table[
    x=SNR_LDPC_5,
    y=BER_LDPC_5,
    col sep=comma
]{data/RM/output_plot128_BER.csv};

% --------------------------------------------------
% LDPC-10
% --------------------------------------------------
\addplot[ldpc10]
table[
    x=SNR_LDPC_10,
    y=BER_LDPC_10,
    col sep=comma
]{data/RM/output_plot128_BER.csv};

% --------------------------------------------------
% LDPC-20
% --------------------------------------------------
\addplot[ldpc20]
table[
    x=SNR_LDPC_20,
    y=BER_LDPC_20,
    col sep=comma
]{data/RM/output_plot128_BER.csv};

% --------------------------------------------------
% AE-SC-8
% --------------------------------------------------
\addplot[aed8]
table[
    x=SNR_AED8,
    y=BER_AED8,
    col sep=comma
]{data/RM/output_plot128_BER.csv};

% --------------------------------------------------
% RM construction + SC
% --------------------------------------------------
\addplot[construction-sc]
table[
    x=SNR_SC_RM,
    y=BER_SC_RM,
    col sep=comma
]{data/RM/output_plot128_BER.csv};
\label{leg:sym-sc}

% --------------------------------------------------
% RM construction + SCL-8
% --------------------------------------------------
\addplot[construction-scl8]
table[
    x=SNR_SCL8_RM,
    y=BER_SCL8_RM,
    col sep=comma
]{data/RM/output_plot128_BER.csv};
\label{leg:sym-scl}

% --------------------------------------------------
% 5G SCL-8 CRC
% --------------------------------------------------
\addplot[5g-scl8-crc]
table[
    x=EbN0_dB,
    y=BER_128_64,
    col sep=comma
]{data/SCL-8-CRC11/data.csv};
\addlegendentry{SCL-8 CRC}
\label{leg:scl8_crc}
\end{axis}

\end{tikzpicture}
    }
    \subfigure{%
        \label{fig:ber_256_rm}
        \begin{tikzpicture}

\begin{axis}[
    xlabel={$E_\mathrm{b}/N_0$ (dB)},
    ylabel={BER},
    width=0.5\linewidth,
    height=0.34\linewidth,
    ymin=1e-8,
    ymax=1e-4,
    xmin=2,
    xmax=6.5,
    clip=true,
    scale=1,
    ymode=log,
    legend style={
        anchor=center,
        cells={anchor=west},
        column sep=2mm,
        thick,
        font=\fontsize{7pt}{7.2pt}\selectfont,
    },
    legend to name=legend:rm_ber_256_163,
    legend columns=3,
    legend pos=north east,
]

% --------------------------------------------------
% FSSC
% --------------------------------------------------
\addplot[fssc]
table[
    x=SNR_5G,
    y=BER_5G,
    col sep=comma
]{data/RM/output_plot256_BER.csv};

% --------------------------------------------------
% AE-DISC-8
% --------------------------------------------------
\addplot[spae8]
table[
    x=SNR_P8,
    y=BER_P8,
    col sep=comma
]{data/RM/output_plot256_BER.csv};

% --------------------------------------------------
% AE-DISC-16
% --------------------------------------------------
\addplot[spae16]
table[
    x=SNR_P16,
    y=BER_P16,
    col sep=comma
]{data/RM/output_plot256_BER.csv};

% --------------------------------------------------
% AE-DISC-32
% --------------------------------------------------
\addplot[spae32]
table[
    x=SNR_P32,
    y=BER_P32,
    col sep=comma
]{data/RM/output_plot256_BER.csv};

% --------------------------------------------------
% FSSCL-8
% --------------------------------------------------
\addplot[scl8]
table[
    x=SNR_SCL8,
    y=BER_SCL8,
    col sep=comma
]{data/RM/output_plot256_BER.csv};

% --------------------------------------------------
% LDPC-5
% --------------------------------------------------
\addplot[ldpc5]
table[
    x=SNR_LDPC_5,
    y=BER_LDPC_5,
    col sep=comma
]{data/RM/output_plot256_BER.csv};

% --------------------------------------------------
% LDPC-10
% --------------------------------------------------
\addplot[ldpc10]
table[
    x=SNR_LDPC_10,
    y=BER_LDPC_10,
    col sep=comma
]{data/RM/output_plot256_BER.csv};

% --------------------------------------------------
% LDPC-20
% --------------------------------------------------
\addplot[ldpc20]
table[
    x=SNR_LDPC_20,
    y=BER_LDPC_20,
    col sep=comma
]{data/RM/output_plot256_BER.csv};

% --------------------------------------------------
% AE-SC-8
% --------------------------------------------------
\addplot[aed8]
table[
    x=SNR_AED8,
    y=BER_AED8,
    col sep=comma
]{data/RM/output_plot256_BER.csv};

% --------------------------------------------------
% [31,57] construction + SC
% --------------------------------------------------
\addplot[construction-sc]
table[
    x=SNR_SC_RM,
    y=BER_SC_RM,
    col sep=comma
]{data/RM/output_plot256_BER.csv};
\label{leg:sym-sc-ber-256}

% --------------------------------------------------
% [31,57] construction + SCL-8
% --------------------------------------------------
\addplot[construction-scl8]
table[
    x=SNR_SCL8_RM,
    y=BER_SCL8_RM,
    col sep=comma
]{data/RM/output_plot256_BER.csv};
\label{leg:sym-scl-ber-256}

%--------------------
%SCL-8 CRC11 5G
%-------
\addplot[5g-scl8-crc]
table[
x=EbN0_dB,
y=BER_256_163,
col sep=comma
]{data/SCL-8-CRC11/data.csv};
\addlegendentry{SCL-8 CRC}

\end{axis}

\end{tikzpicture}
    }

    \RMDecoderLegend

    \caption{\ac{BER} and \ac{FER} performance comparison of the proposed \ac{AE-DISC} decoder with other decoders for RM codes and the 5G LDPC code.}
    \label{fig:fer_ber_rm}
\end{figure*}

This section presents the simulation setup and evaluates the \ac{FER} and
\ac{BER} performance, complexity, and latency of the proposed decoder. In addition, the mismatch rate
between \ac{AE-DISC} and \ac{AE-SC} is analyzed, and the impact of fixed-point
quantization on decoding performance is investigated.

\subsection{Simulation Setup}
Simulations are conducted over an \ac{AWGN} channel with \ac{BPSK} modulation, i.e., the transmitted signal is given by $\boldsymbol{s} = \boldsymbol{1} - 2\boldsymbol{x}$, and the received vector is
\begin{align}
\boldsymbol{y} = \boldsymbol{s} + \boldsymbol{n}, \qquad \boldsymbol{n} \sim \mathcal{N}(\boldsymbol{0}, \sigma^2 \boldsymbol{I}),
\end{align}
where $\boldsymbol{n}$ is zero-mean Gaussian noise with variance $\sigma^2$.  

Polar codes for \ac{AE-DISC} and \ac{AE-SC} are constructed from $\mathcal{I}_\mathrm{min} = \{23, 112\}$ for $\mathcal{P}(128, 64)$ and $\mathcal{I}_\mathrm{min} = \{31, 57\}$ for $\mathcal{P}(256, 128)$, as described in ~\Cref{subsec:aed}. 
Moreover, we consider the \(\mathrm{RM}(3,7)\) and \(\mathrm{RM}(4,8)\) codes with parameters $(128,64)$ and $(256,163)$, respectively. The permutations are randomly sampled from the respective \ac{BLTA} groups of the Polar codes and general affine groups for \ac{RM} codes.

As baselines, we evaluate 5G \ac{LDPC} codes with the \ac{LMS} decoder from~\cite{sionna}, as well as Polar codes constructed with the 5G Polar sequence~\cite{3gpp2018multiplexing} under \ac{SC} and \ac{SCL}-8 (with and without 5G NR 11-bit \ac{CRC}). In addition, \ac{SC} and \ac{SCL}-8 results are shown for the same symmetric Polar code design used with \ac{AE-DISC}. The \ac{PPV} bound for the same block length is shown~\cite{polyanskiy2010channel}. The evaluation focuses on \ac{FER} values between $10^{-5}$ and $10^{-7}$, which corresponds to the target operating region of \ac{HRLLC}.

\subsection{FER and BER Performance}
\label{subsec:ber_fer_performance}
The \ac{FER} performance comparison in~\Cref{fig:polar_fer} between Polar and \ac{LDPC} decoders shows that, for \( \mathcal{P}(128, 64) \), \ac{AE-DISC} consistently outperforms both \ac{FSSC} and \ac{LDPC} \ac{LMS} decoders across the target \ac{FER} range of \( 10^{-5} \) to \( 10^{-7} \). \ac{AE-DISC}-8 achieves a 0.8\,dB gain over \ac{LDPC} \ac{LMS} with 5 iterations and a 0.3\,dB gain over the 5G \ac{FSSC} baseline, while matching the performance of \ac{LDPC} \ac{LMS} with 20 iterations at \ac{FER} of \(10^{-6}\). 
Increasing the ensemble size from 8 to 16 and 32 further improves \ac{FER} performance.
Compared to \ac{AE-SC}, the weaker constituent decoder of \ac{AE-DISC} generally leads to larger ensemble sizes required to match the \ac{FER}. 
However, we do not notice a weaker limiting performance, and \ac{AE-DISC}-32 achieves nearly identical performance to \ac{AE-SC}-8.  

For \( \mathcal{P}(256, 128) \), the results demonstrate similar trends. \ac{AE-DISC}-8 achieves a 0.8\,dB coding gain over \ac{FSSC} at an \ac{FER} of \( 10^{-6} \). Expanding the ensemble to 16 and 32 yields gains of 1.1\,dB and 1.4\,dB, respectively, compared to the 5G \ac{FSSC} decoder at the same \ac{FER}. The \ac{AE-DISC}-8 curve achieves 0.5 dB gain over the \ac{LDPC} decoder with 5 iterations, while \ac{AE-DISC}-16 surpasses the \ac{LDPC} 10-iteration performance by 0.2\,dB. Again, \ac{AE-DISC}-32 exhibits similar \ac{FER} performance to \ac{AE-SC}-8 at \ac{FER} of \( 10^{-6} \).

The BER performance shows that increasing the ensemble size consistently improves the error-correction capability of \ac{AE-DISC}. For both $\mathcal{P}(128,64)$ and $\mathcal{P}(256,128)$, \ac{AE-DISC}-32 approaches the \ac{BER} performance of \ac{AE-SC}-8 at high \ac{SNR} region while retaining the lower-latency advantage of the \ac{DISC} constituent decoder. 

For the \(\mathrm{RM}(3,7)\) and \(\mathrm{RM}(4,8)\) codes, \Cref{fig:fer_ber_rm} compares the \ac{FER} performance of the proposed \ac{AE-DISC} decoder with \ac{FSSC}, \ac{SCL}-8, \ac{AE-SC}-8, and \ac{LDPC} decoders for identical code parameters. Again, for both codes, the performance of \ac{AE-DISC} improves consistently as the ensemble size increases from 8 to 32, and no error floor is observed within the simulated error-rate range. %
At \ac{FER} of approximately $10^{-6}$, \ac{AE-DISC}-32 provides $1.2\,\mathrm{dB}$ gain over the conventional \ac{FSSC} decoder. Compared to the \ac{SCL}-8 decoder, \ac{AE-DISC}-32 provides $0.1\,\mathrm{dB}$ gain for the \(\mathrm{RM}(3,7)\) code and $0.2\,\mathrm{dB}$ gain for the and \(\mathrm{RM}(4,8)\) code. 

\ac{AE-DISC}-32 is also within $0.2\ \mathrm{dB}$ gap from \ac{AE-SC}-8 \ac{FER} for both \ac{RM} codes. For the and \(\mathrm{RM}(4,8)\)  code, \ac{AE-DISC}-32 achieves better \ac{FER} performance than \ac{LDPC}-20. \Ac{AE-DISC}-16 achieves performance close to \ac{SCL}-8, while \ac{AE-DISC}-32 consistently outperforms \ac{SCL}-8 and approaches the performance of \ac{AE-SC}-8.

\subsection{\ac{AE-DISC} and \ac{AE-SC} Mismatch Rate}

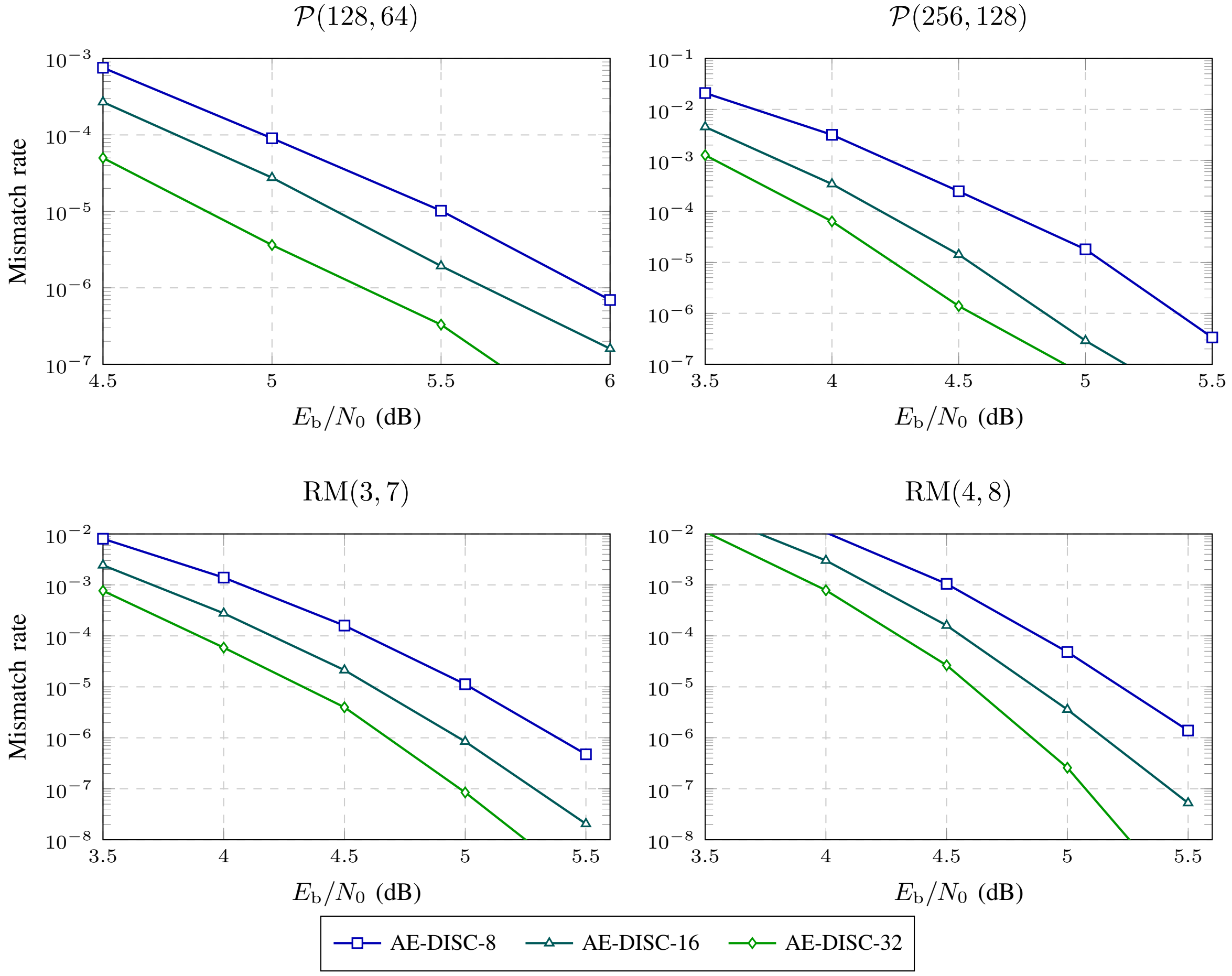
\begin{figure*}[t]
    \centering
    \begin{tikzpicture}

% \pgfplotsset{
%     label style={
%         font=\fontsize{9pt}{7.2}\selectfont
%     },
%     tick label style={
%         font=\fontsize{7pt}{7.2}\selectfont
%     }
% }

% =========================================================
% Polar row
% =========================================================
\begin{scope}

\begin{groupplot}[
    group style={
        group size=2 by 1,
        horizontal sep=1.2cm,
        group name=polarplots
    },
    width=0.44\textwidth,
    height=0.30\textwidth,
    xlabel={$E_\mathrm{b}/N_0$ (dB)},
    ylabel={Mismatch rate},
    ymode=log,
    grid=major,
    grid style={dashed,gray!40},
    clip=true,
    unbounded coords=discard,
    filter discard warning=false
]

% ---------------------------------------------------------
% Polar (128,64)
% ---------------------------------------------------------
\nextgroupplot[
    title={$\mathcal{P}(128,64)$},
    xmin=4.5,
    xmax=6,
    ymin=1e-7,
    ymax=1e-3,
    xtick={4.5,5,5.5,6,6.5},
    legend to name=commonlegend
]

% AE-DISC-8: rows 0-4
\addplot[
    spae8
]
table[
    col sep=comma,
    x=ebn0_db,
    y=spae_aed_full_mismatch_rate,
    restrict expr to domain={\coordindex}{0:4}
]{data/failure_rate_aed/polar.csv};
\label{leg:mismatch-aedisc-8}

% AE-DISC-16: rows 5-8
\addplot[
    spae16
]
table[
    col sep=comma,
    x=ebn0_db,
    y=spae_aed_full_mismatch_rate,
    restrict expr to domain={\coordindex}{5:8}
]{data/failure_rate_aed/polar.csv};
\label{leg:mismatch-aedisc-16}

% AE-DISC-32: rows 9-12
\addplot[
    spae32
]
table[
    col sep=comma,
    x=ebn0_db,
    y=spae_aed_full_mismatch_rate,
    restrict expr to domain={\coordindex}{9:12}
]{data/failure_rate_aed/polar.csv};
\label{leg:mismatch-aedisc-32}

% ---------------------------------------------------------
% Polar (256,128)
% ---------------------------------------------------------

\nextgroupplot[
    title={$\mathcal{P}(256,128)$},
    ylabel={},
    xmin=3.5,
    xmax=5.5,
    ymin=1e-7,
    ymax=1e-1,
    xtick={3.5,4,4.5,5,5.5},
    ytick={
        1e-8,1e-7,1e-6,1e-5,
        1e-4,1e-3,1e-2,1e-1
    },
    grid=major,
    grid style={dashed,gray!40},
    ymajorgrids=true
]

% AE-DISC-8: rows 13-18
\addplot[
    spae8
]
table[
    col sep=comma,
    x=ebn0_db,
    y=spae_aed_full_mismatch_rate,
    restrict expr to domain={\coordindex}{13:18}
]{data/failure_rate_aed/polar.csv};

% AE-DISC-16: rows 19-23
\addplot[
    spae16
]
table[
    col sep=comma,
    x=ebn0_db,
    y=spae_aed_full_mismatch_rate,
    restrict expr to domain={\coordindex}{19:23}
]{data/failure_rate_aed/polar.csv};

% AE-DISC-32: rows 24-27
\addplot[
    spae32
]
table[
    col sep=comma,
    x=ebn0_db,
    y=spae_aed_full_mismatch_rate,
    restrict expr to domain={\coordindex}{24:27}
]{data/failure_rate_aed/polar.csv};

\end{groupplot}

\end{scope}

% =========================================================
% RM row -- UPDATED DATA
% =========================================================
\begin{scope}[yshift=-6cm]

\begin{groupplot}[
    group style={
        group size=2 by 1,
        horizontal sep=1.2cm,
        group name=rmplots
    },
    width=0.44\textwidth,
    height=0.30\textwidth,
    xlabel={$E_\mathrm{b}/N_0$ (dB)},
    ylabel={Mismatch rate},
    ymode=log,
    grid=major,
    grid style={dashed,gray!40},
    ymajorgrids=true,
    clip=true,
    unbounded coords=discard,
    filter discard warning=false
]

% ---------------------------------------------------------
% RM(3,7) (128,64)
% ---------------------------------------------------------
\nextgroupplot[
    title={$\mathrm{RM}(3,7)$},
    xmin=3.5,
    xmax=5.6,
    ymin=1e-8,
    ymax=1e-2,
    xtick={3.5,4,4.5,5,5.5},
    ytick={
        1e-10,1e-9,1e-8,1e-7,
        1e-6,1e-5,1e-4,1e-3,1e-2
    },
    ymajorgrids=true
]

% AE-DISC-8: rows 0-4
\addplot[
    spae8
]
table[
    col sep=comma,
    x=ebn0_db,
    y=spae_aed_full_mismatch_rate,
    restrict expr to domain={\coordindex}{0:4}
]{data/failure_rate_aed/rm.csv};

% AE-DISC-16: rows 5-9
\addplot[
    spae16
]
table[
    col sep=comma,
    x=ebn0_db,
    y=spae_aed_full_mismatch_rate,
    restrict expr to domain={\coordindex}{5:9}
]{data/failure_rate_aed/rm.csv};

% AE-DISC-32: rows 10-14
\addplot[
    spae32
]
table[
    col sep=comma,
    x=ebn0_db,
    y=spae_aed_full_mismatch_rate,
    restrict expr to domain={\coordindex}{10:14}
]{data/failure_rate_aed/rm.csv};

% ---------------------------------------------------------
% RM(4,8) (256,163)
% ---------------------------------------------------------
\nextgroupplot[
    title={$\mathrm{RM}(4,8)$},
    ylabel={},
    xmin=3.5,
    xmax=5.6,
    ymin=1e-8,
    ymax=1e-2,
    xtick={3.5,4,4.5,5,5.5},
    ytick={
        1e-10,1e-9,1e-8,1e-7,
        1e-6,1e-5,1e-4,1e-3,
        1e-2,1e-1
    },
    ymajorgrids=true
]

% AE-DISC-8: rows 15-19
\addplot[
    spae8
]
table[
    col sep=comma,
    x=ebn0_db,
    y=spae_aed_full_mismatch_rate,
    restrict expr to domain={\coordindex}{15:19}
]{data/failure_rate_aed/rm.csv};

% AE-DISC-16: rows 20-24
\addplot[
    spae16
]
table[
    col sep=comma,
    x=ebn0_db,
    y=spae_aed_full_mismatch_rate,
    restrict expr to domain={\coordindex}{20:24}
]{data/failure_rate_aed/rm.csv};

% AE-DISC-32: rows 25-29
\addplot[
    spae32
]
table[
    col sep=comma,
    x=ebn0_db,
    y=spae_aed_full_mismatch_rate,
    restrict expr to domain={\coordindex}{25:29}
]{data/failure_rate_aed/rm.csv};

\end{groupplot}

\end{scope}

% =========================================================
% Shared legend for all four panels
% % =========================================================
% \node[
%     anchor=north
% ]
% at ($(rmplots c1r1.south)!0.5!(rmplots c2r1.south)+(0,-1.25cm)$)
% {\pgfplotslegendfromname{commonlegend}};

\end{tikzpicture}
    \MismatchRateLegend
    \caption{Mismatch rate between \ac{AE-DISC} and \ac{AE-SC} versus
    $E_\mathrm{b}/N_0$ for Polar and \ac{RM} codes. }
    \label{fig:mismatch_rate}
\end{figure*}

To compare the proposed \ac{AE-DISC} and conventional \ac{AE-SC} decoding, we define the \emph{mismatch rate} as the fraction of simulated frames for which the final decoded codeword returned by \ac{AE-DISC} differs from that returned by \ac{AE-SC} for the same received frame given the same permutation set.
\Cref{fig:mismatch_rate} shows the mismatch rate for the considered codes. For both $\mathcal{P}(128,64)$ and $\mathcal{P}(256,128)$, the mismatch rate between \ac{AE-DISC} and \ac{AE-SC} decreases rapidly with increasing $E_\mathrm{b}/N_0$ and with increasing ensemble size. In particular, \ac{AE-DISC}-32 consistently achieves the lowest mismatch rate. This indicates that, with a sufficiently large candidate set, \ac{AE-DISC} can closely reproduce the candidate selection of \ac{AE-SC}.
A similar trend is observed for \ac{RM} codes, i.e., higher \ac{SNR} leads to lower mismatch rate.

\subsection{Quantization Scheme}
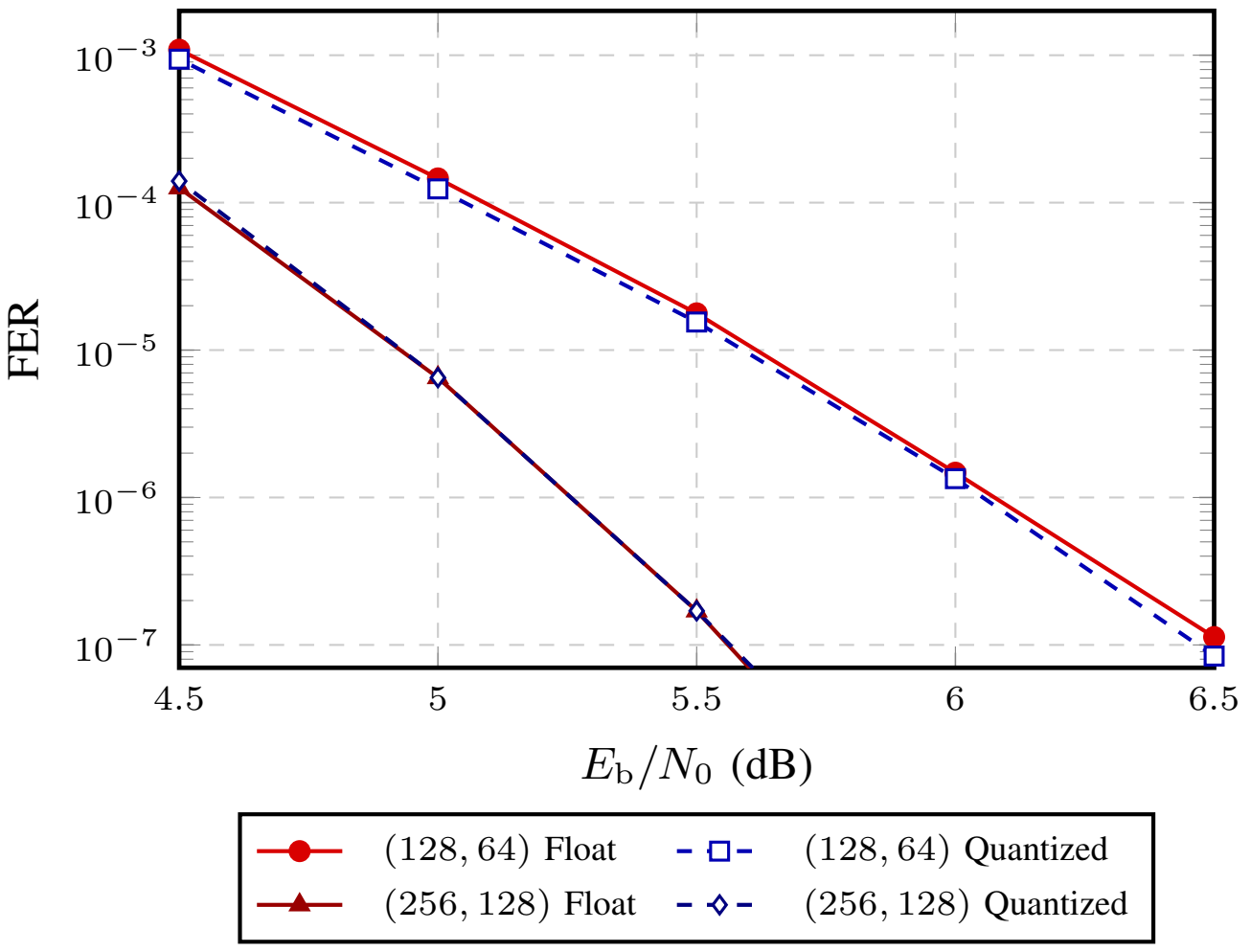
\begin{figure}[t]
\centering
\begin{tikzpicture}

% \pgfplotsset{
%     label style = {font=\fontsize{9pt}{7.2}\selectfont},
%     tick label style = {font=\fontsize{7pt}{7.2}\selectfont}
% }

\begin{axis}[
    xlabel={$E_\mathrm{b}/N_0$ (dB)},
    ylabel={FER},
    grid=major,
    grid style={dashed,gray!40},
    width=\linewidth,
    height=0.7\linewidth,
    line width=0.9pt,
    mark size=1.8pt,
    % ticklabel style={font=\footnotesize},
    % label style={font=\footnotesize},
    ymin=7e-8,
    ymax=2e-3,
    xmin=4.5,
    xmax=6.5,
    clip=true,
    ymode=log,
legend style={
    at={(0.5,-0.22)},
    anchor=north,
    cells={anchor=west},
    column sep=2mm,
    thick,
    font=\fontsize{7pt}{8.4pt}\selectfont,
},
legend columns=2
]

% ======================================================
% N = 128
% ======================================================

\addplot[AEDFP128]
table[
    x=SNR_unquan,
    y=FER_unquan,
    col sep=comma
]
{data/SPAE_Quan_unquan_128/128-64.csv};
\addlegendentry{$(128,64)$ Float}

\addplot[AEDQ128]
table[
    x=SNR_quan,
    y=FER_quan,
    col sep=comma
]
{data/SPAE_Quan_unquan_128/128-64.csv};
\addlegendentry{$(128,64)$ Quantized}

% ======================================================
% N = 256
% ======================================================

\addplot[AEDFP256]
table[
    x=SNR_unquan,
    y=FER_unquan,
    col sep=comma
]
{data/SPAE_Quan_unquan_256/256-128.csv};
\addlegendentry{$(256,128)$ Float}

\addplot[AEDQ256]
table[
    x=SNR_quan,
    y=FER_quan,
    col sep=comma
]
{data/SPAE_Quan_unquan_256/256-128.csv};
\addlegendentry{$(256,128)$ Quantized}

\end{axis}
\end{tikzpicture}
\caption{\ac{FER} performance comparison between floating-point and quantized \ac{AE-DISC}-8 decoding for $\mathcal{P}(128,64)$ and $\mathcal{P}(256,128)$.}
\label{fig:quantization}
\end{figure}
Lower-precision quantization is a key enabler for achieving high throughput and low latency in hardware implementations. An important property of Polar codes is that \ac{SC} decoding remains effective under low-precision quantization, typically in the range of 4 to 6 bits \cite{tong2023fast_latency}. 
For ensemble decoders, different quantization formats are considered for \ac{LLR} quantization and Euclidean distance quantization.

A signed fixed-point representation in two's complement format is used for \ac{LLR} quantization, with a total word length of 5 bits including 1 fractional bit. The Euclidean distance metric for deciding the best codeword is represented using 6 bits with 2 fractional bits. As shown in \cref{fig:quantization}, using these proposed parameters, the performance loss compared to the floating-point implementation is negligible.

\subsection{Computational Complexity}
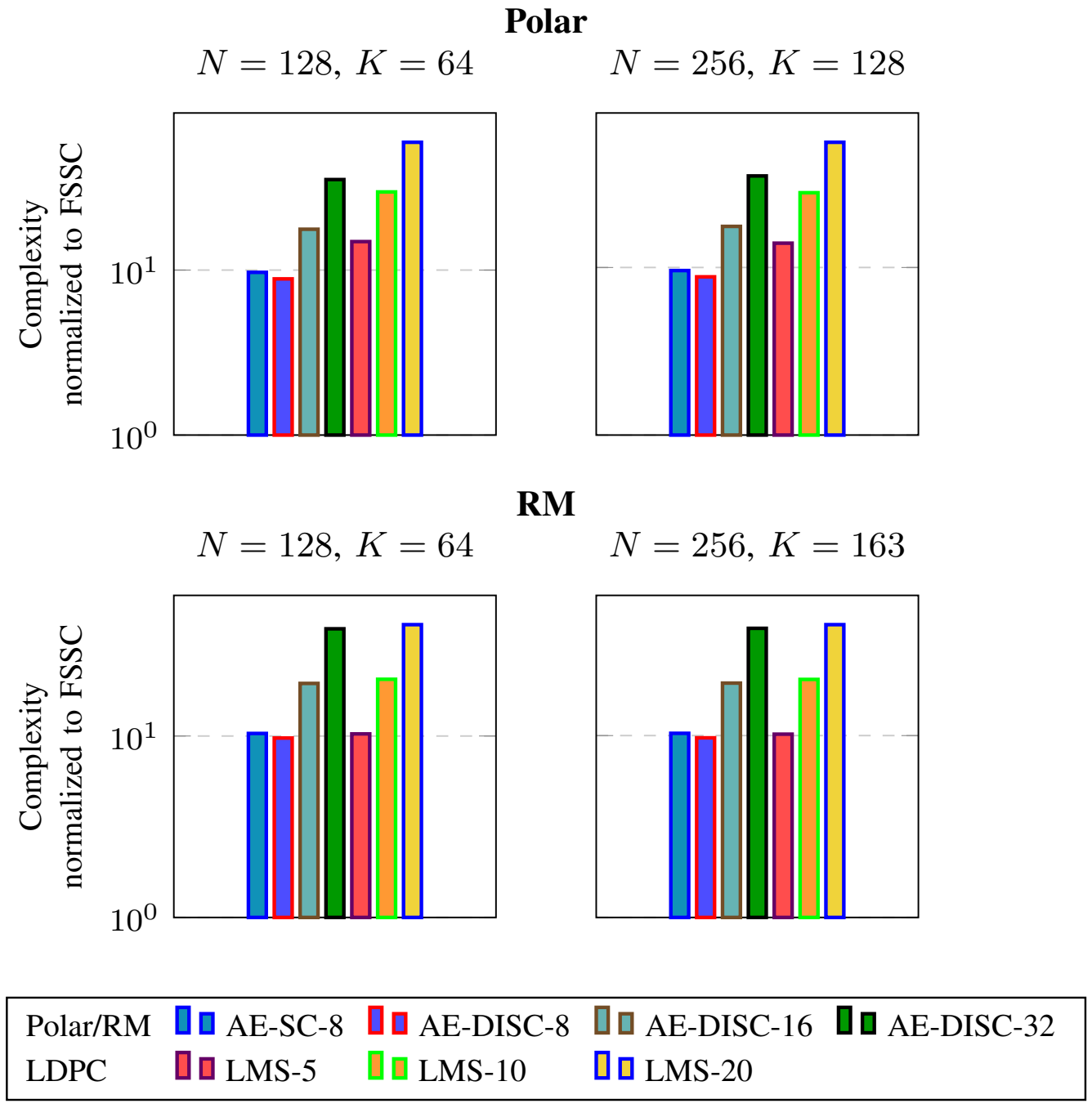
\begin{figure}
    \centering
    % \pgfplotsset{compat=1.18}

\begin{tikzpicture}

\begin{groupplot}[
    group style={
        group size=2 by 2,
        horizontal sep=0.10\columnwidth,
        vertical sep=0.16\columnwidth,
    },
    ybar,
    width=0.5\columnwidth,
    height=0.5\columnwidth,
    ymin=1,
    ymode=log,
    xtick=\empty,
    xtick style={draw=none},
    ticklabel style={font=\footnotesize},
    label style={font=\footnotesize},
    title style={
        font=\small,
        yshift=-1pt
    },
    yticklabel style={font=\footnotesize},
    grid=major,
    grid style={dashed,gray!40},
    enlarge x limits=0.5,
    clip=false,
]

% ============================================================
% (a) Polar: N = 128, K = 64
% ============================================================
\nextgroupplot[
    ylabel={\shortstack{Complexity \\normalized to FSSC}},
    title={$N=128$, $K=64$},
]

\addplot+[
    fill=cyan!70!black,
    /pgf/bar width=4.5pt
] table[col sep=comma,
    x=N,
    y=AED-8_N,
    restrict expr to domain={\thisrow{N}}{128:128}
]{data/complexity.csv};
\label{bar:aed8}

\addplot+[
    fill=blue!70,
    /pgf/bar width=4.5pt
] table[col sep=comma,
    x=N,
    y=SPAE-8_N,
    restrict expr to domain={\thisrow{N}}{128:128}
]{data/complexity.csv};
\label{bar:spae8}

\addplot+[
    fill=teal!60,
    /pgf/bar width=4.5pt
] table[col sep=comma,
    x=N,
    y=SPAE-16_N,
    restrict expr to domain={\thisrow{N}}{128:128}
]{data/complexity.csv};
\label{bar:spae16}

\addplot+[
    fill=green!60!black,
    /pgf/bar width=4.5pt
] table[col sep=comma,
    x=N,
    y=SPAE-32_N,
    restrict expr to domain={\thisrow{N}}{128:128}
]{data/complexity.csv};
\label{bar:spae32}

\addplot+[
    fill=red!70,
    /pgf/bar width=4.5pt
] table[col sep=comma,
    x=N,
    y=LDPC_5_N,
    restrict expr to domain={\thisrow{N}}{128:128}
]{data/complexity.csv};
\label{bar:lms5}

\addplot+[
    fill=orange!80,
    /pgf/bar width=4.5pt
] table[col sep=comma,
    x=N,
    y=LDPC_10_N,
    restrict expr to domain={\thisrow{N}}{128:128}
]{data/complexity.csv};
\label{bar:lms10}

\addplot+[
    fill=yellow!70!brown,
    /pgf/bar width=4.5pt
] table[col sep=comma,
    x=N,
    y=LDPC_20_N,
    restrict expr to domain={\thisrow{N}}{128:128}
]{data/complexity.csv};
\label{bar:lms20}

% ============================================================
% (b) Polar: N = 256, K = 128
% ============================================================
\nextgroupplot[
    title={$N=256$, $K=128$},
    yticklabels=\empty,
]

\addplot+[
    fill=cyan!70!black,
    /pgf/bar width=4.5pt
] table[col sep=comma,
    x=N,
    y=AED-8_N,
    restrict expr to domain={\thisrow{N}}{256:256}
]{data/complexity.csv};

\addplot+[
    fill=blue!70,
    /pgf/bar width=4.5pt
] table[col sep=comma,
    x=N,
    y=SPAE-8_N,
    restrict expr to domain={\thisrow{N}}{256:256}
]{data/complexity.csv};

\addplot+[
    fill=teal!60,
    /pgf/bar width=4.5pt
] table[col sep=comma,
    x=N,
    y=SPAE-16_N,
    restrict expr to domain={\thisrow{N}}{256:256}
]{data/complexity.csv};

\addplot+[
    fill=green!60!black,
    /pgf/bar width=4.5pt
] table[col sep=comma,
    x=N,
    y=SPAE-32_N,
    restrict expr to domain={\thisrow{N}}{256:256}
]{data/complexity.csv};

\addplot+[
    fill=red!70,
    /pgf/bar width=4.5pt
] table[col sep=comma,
    x=N,
    y=LDPC_5_N,
    restrict expr to domain={\thisrow{N}}{256:256}
]{data/complexity.csv};

\addplot+[
    fill=orange!80,
    /pgf/bar width=4.5pt
] table[col sep=comma,
    x=N,
    y=LDPC_10_N,
    restrict expr to domain={\thisrow{N}}{256:256}
]{data/complexity.csv};

\addplot+[
    fill=yellow!70!brown,
    /pgf/bar width=4.5pt
] table[col sep=comma,
    x=N,
    y=LDPC_20_N,
    restrict expr to domain={\thisrow{N}}{256:256}
]{data/complexity.csv};

% ============================================================
% (c) RM: N = 128, K = 64
% ============================================================
\nextgroupplot[
    ylabel={\shortstack{Complexity \\normalized to FSSC}},
    title={$N=128$, $K=64$},
]

\addplot+[
    fill=cyan!70!black,
    /pgf/bar width=4.5pt
] table[
    col sep=comma,
    x=N,
    y=AED-8_N,
    restrict expr to domain={\thisrow{N}}{128:128}
]{data/RM/complexity.csv};

\addplot+[
    fill=blue!70,
    /pgf/bar width=4.5pt
] table[
    col sep=comma,
    x=N,
    y=SPAE-8_N,
    restrict expr to domain={\thisrow{N}}{128:128}
]{data/RM/complexity.csv};

\addplot+[
    fill=teal!60,
    /pgf/bar width=4.5pt
] table[
    col sep=comma,
    x=N,
    y=SPAE-16_N,
    restrict expr to domain={\thisrow{N}}{128:128}
]{data/RM/complexity.csv};

\addplot+[
    fill=green!60!black,
    /pgf/bar width=4.5pt
] table[
    col sep=comma,
    x=N,
    y=SPAE-32_N,
    restrict expr to domain={\thisrow{N}}{128:128}
]{data/RM/complexity.csv};

\addplot+[
    fill=red!70,
    /pgf/bar width=4.5pt
] table[
    col sep=comma,
    x=N,
    y=LDPC_5_N,
    restrict expr to domain={\thisrow{N}}{128:128}
]{data/RM/complexity.csv};

\addplot+[
    fill=orange!80,
    /pgf/bar width=4.5pt
] table[
    col sep=comma,
    x=N,
    y=LDPC_10_N,
    restrict expr to domain={\thisrow{N}}{128:128}
]{data/RM/complexity.csv};

\addplot+[
    fill=yellow!70!brown,
    /pgf/bar width=4.5pt
] table[
    col sep=comma,
    x=N,
    y=LDPC_20_N,
    restrict expr to domain={\thisrow{N}}{128:128}
]{data/RM/complexity.csv};

% ============================================================
% (d) RM: N = 256, K = 163
% ============================================================
\nextgroupplot[
    title={$N=256$, $K=163$},
    yticklabels=\empty,
]

\addplot+[
    fill=cyan!70!black,
    /pgf/bar width=4.5pt
] table[
    col sep=comma,
    x=N,
    y=AED-8_N,
    restrict expr to domain={\thisrow{N}}{256:256}
]{data/RM/complexity.csv};

\addplot+[
    fill=blue!70,
    /pgf/bar width=4.5pt
] table[
    col sep=comma,
    x=N,
    y=SPAE-8_N,
    restrict expr to domain={\thisrow{N}}{256:256}
]{data/RM/complexity.csv};

\addplot+[
    fill=teal!60,
    /pgf/bar width=4.5pt
] table[
    col sep=comma,
    x=N,
    y=SPAE-16_N,
    restrict expr to domain={\thisrow{N}}{256:256}
]{data/RM/complexity.csv};

\addplot+[
    fill=green!60!black,
    /pgf/bar width=4.5pt
] table[
    col sep=comma,
    x=N,
    y=SPAE-32_N,
    restrict expr to domain={\thisrow{N}}{256:256}
]{data/RM/complexity.csv};

\addplot+[
    fill=red!70,
    /pgf/bar width=4.5pt
] table[
    col sep=comma,
    x=N,
    y=LDPC_5_N,
    restrict expr to domain={\thisrow{N}}{256:256}
]{data/RM/complexity.csv};

\addplot+[
    fill=orange!80,
    /pgf/bar width=4.5pt
] table[
    col sep=comma,
    x=N,
    y=LDPC_10_N,
    restrict expr to domain={\thisrow{N}}{256:256}
]{data/RM/complexity.csv};

\addplot+[
    fill=yellow!70!brown,
    /pgf/bar width=4.5pt
] table[
    col sep=comma,
    x=N,
    y=LDPC_20_N,
    restrict expr to domain={\thisrow{N}}{256:256}
]{data/RM/complexity.csv};

\end{groupplot}

% ============================================================
% Row labels
% ============================================================

% Polar row label
\node[
    anchor=south,
    font=\small\bfseries
] at (
    $(group c1r1.north west)!0.5!(group c2r1.north east)
    +(0,0.58cm)$
) {Polar};

% RM row label
\node[
    anchor=south,
    font=\small\bfseries
] at (
    $(group c1r2.north west)!0.5!(group c2r2.north east)
    +(0,0.58cm)$
) {RM};

% ============================================================
% Common legend below all four plots
% ============================================================

\node[
    draw=black,
    line width=0.5pt,
    align=left,
    anchor=north,
    font=\footnotesize,
    inner sep=2pt,
] at (
    $(group c1r2.south west)!0.5!(group c2r2.south east)
    +(0,-0.70cm)$
) {
    \setlength{\tabcolsep}{3pt}
    \begin{tabular}{@{}lllll@{}}
        Polar/RM
        & \ref{bar:aed8} AE-SC-8
        & \ref{bar:spae8} AE-DISC-8
        & \ref{bar:spae16} AE-DISC-16
        & \ref{bar:spae32} AE-DISC-32
        \\[1pt]
        LDPC
        & \ref{bar:lms5} LMS-5
        & \ref{bar:lms10} LMS-10
        & \ref{bar:lms20} LMS-20
        &
    \end{tabular}
};

\end{tikzpicture}
    \caption{Decoding complexity in terms of \ac{LLR} operations for different decoders, normalized to the corresponding \ac{FSSC} baseline.
    }
\label{fig:complexity_sidebyside}
\end{figure}

\Cref{fig:complexity_sidebyside} compares the computational complexity of \ac{FSSC}, \ac{AE-SC}, \ac{AE-DISC} and \ac{LDPC} \ac{LMS} decoding in terms of \ac{LLR} operations, as introduced in \Cref{ssec:prelim_complexity} and \Cref{ssec:aedisc_complexity}. 
For all code families, the number
of operations is normalized to that of the corresponding \ac{FSSC}
decoder with the same parameters $N$ and $K$.

Across both Polar and RM codes, the computational complexity of
\ac{AE-DISC} increases approximately linearly with the ensemble size.
For a given ensemble size, \ac{AE-DISC} requires slightly fewer
operations than the corresponding \ac{AE-SC} decoder,
because each \ac{DISC} decoding path eliminates $N/2$ additions.
The complexity of the \ac{LDPC} \ac{LMS} decoder
increases approximately linearly with the number of iterations.

For both Polar and RM codes, increasing the ensemble size from \ac{AE-DISC}-8 to \ac{AE-DISC}-32 improves the error-rate performance
at the cost of an approximately linear increase in complexity.

At similar \ac{FER}, \ac{AE-DISC} requires fewer operations than \ac{LMS} decoding; for example, for $\mathcal{P}(128,64)$, \ac{AE-DISC}-8 matches the \ac{FER} of \ac{LMS}-20 at $10^{-6}$ with one-sixth of its operations, and for $\mathcal{P}(256,128)$, \ac{AE-DISC}-16 outperforms \ac{LMS}-10 by 0.2\,dB with two-thirds of its complexity.

For $\mathrm{RM}(3,7)$, \ac{AE-DISC}-8 outperforms \ac{LMS}-20 with one-quarter of its complexity, whereas for $\mathrm{RM}(4,8)$, \ac{AE-DISC}-8, \ac{AE-DISC}-16, and \ac{AE-DISC}-32 achieve \ac{FER} comparable to \ac{LMS}-5, \ac{LMS}-10, and \ac{LMS}-20, respectively. These results demonstrate a favorable complexity--error-correction performance balance for \ac{AE-DISC} across both Polar and RM codes.
\subsection{Latency--Reliability Performance}

\begin{table}[t]
\centering
\def\arraystretch{1.35}
\caption{Latency model of fast nodes in \ac{SCL} decoding.}
\begingroup
\begin{tabular}{@{}cc@{}}
\hline
Node & Time steps $T$ \\ \hline

Rate-0 & $1$ \\
REP    & $2$ \\
Rate-1 & $\min(L-1,\,N_v)$ \\
SPC    & $\min(L,\,N_v)+1$ \\
FHT, $\mathrm{RM}(1,m)$ & $2m$ \\

\hline
\end{tabular}
\endgroup
\label{tab:scl_latency}
\end{table}

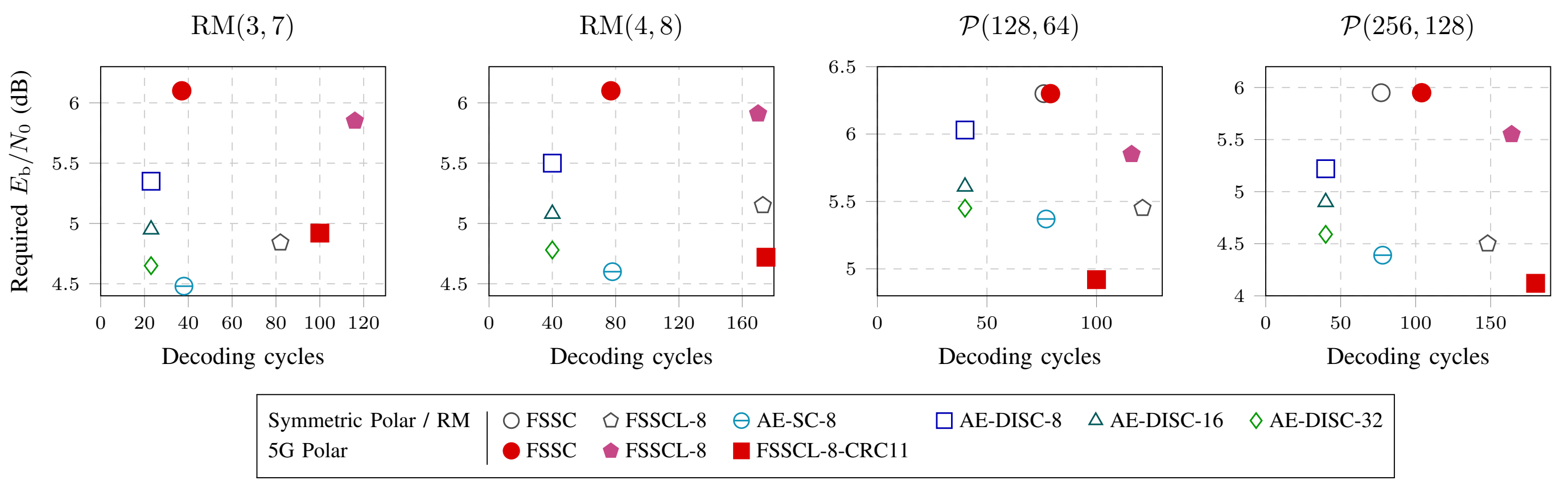
\begin{figure*}[t]
        \centering
        \resizebox{\linewidth}{!}{\begin{tikzpicture}

% =========================================================
% Select one row from the CSV.
% Row indexing begins at zero after skipping the header.
% =========================================================
\pgfplotsset{
    select row/.style={
        x filter/.code={
            \ifnum\coordindex=#1\relax
            \else
                \def\pgfmathresult{}
            \fi
        }
    }
}

% =========================================================
% CSV file
% =========================================================

% =========================================================
% Plot one CSV row.
%
% #1 = zero-based row index after the CSV header
% #2 = plot style defined in plotstyle.tex
% =========================================================
\newcommand{\plotpoint}[2]{%
    \addplot[
        #2,
        only marks,
        forget plot,
        mark size=3.2pt,
        select row=#1
    ]
    table[
        header=false,
        skip first n=1,
        x index=2,
        y index=3,
        col sep=comma
    ]{data/pareto/latency_snr.csv};
}

% =========================================================
% Four-panel plot
% =========================================================
\begin{groupplot}[
    group style={
        group size=4 by 1,
        horizontal sep=1.35cm
    },
    scale only axis,
    width=0.205\linewidth,
    height=0.165\linewidth,
    xlabel={Decoding cycles},
    ylabel={Required $E_\mathrm{b}/N_0$ (dB)},
    unbounded coords=discard,
    tick align=outside,
    tick pos=left
]

% =========================================================
% RM(3,7), (128,64)
%
% RM rows:                    0--5
% 5G FSSCL-8 (128,64):        14
% 5G FSSCL-8-CRC11 (128,64):  15
% =========================================================
\nextgroupplot[
    xmin=0,
    xmax=130,
    ymin=4.4,
    ymax=6.3,
    xtick={0,20,40,60,80,100,120},
    title={$\mathrm{RM}(3,7)$}
]

\plotpoint{0}{fssc}                  % GMC: (37,6.10)
\plotpoint{2}{spae8}                 % AE-DISC-8: (23,5.35)
\plotpoint{3}{spae16}                % AE-DISC-16: (23,4.95)
\plotpoint{4}{spae32}                % AE-DISC-32: (23,4.65)
\plotpoint{5}{aed8}                  % AE-SC-8: (38,4.48)

% RM Fast-SCL-8 (same open-pentagon style as Sym. FSSCL-8)
\plotpoint{1}{construction-scl8}     % (82,4.84)

% 5G Fast-SSCL-SPC-8 for (128,64)
\plotpoint{14}{scl8}                 % (116,5.85)
\plotpoint{15}{5g-scl8-crc}              % CRC11: (100,4.92)

% =========================================================
% RM(4,8), (256,163)
%
% RM rows:                     6--11
% 5G FSSCL-8 (256,163):        30
% 5G FSSCL-8-CRC11 (256,163):  31
% =========================================================
\nextgroupplot[
    xmin=0,
    xmax=180,
    ymin=4.4,
    ymax=6.3,
    xtick={0,40,80,120,160},
    ylabel={},
    title={$\mathrm{RM}(4,8)$}
]

\plotpoint{6}{fssc}                  % GMC: (77,6.10)
\plotpoint{8}{spae8}                 % AE-DISC-8: (40,5.50)
\plotpoint{9}{spae16}                % AE-DISC-16: (40,5.08)
\plotpoint{10}{spae32}               % AE-DISC-32: (40,4.78)
\plotpoint{11}{aed8}                 % AE-SC-8: (78,4.60)

% RM Fast-SCL-8 (same open-pentagon style as Sym. FSSCL-8)
\plotpoint{7}{construction-scl8}     % (173,5.15)

% 5G Fast-SSCL-SPC-8 for (256,163)
\plotpoint{30}{scl8}                 % (170,5.91)
\plotpoint{31}{5g-scl8-crc}              % CRC11: (175,4.72)

% =========================================================
% Polar (128,64)
% Rows 12--20
% =========================================================
\nextgroupplot[
    xmin=0,
    xmax=130,
    ymin=4.8,
    ymax=6.5,
    ylabel={},
    title={$\mathcal{P}(128,64)$}
]

% UPO/symmetric construction
\plotpoint{12}{construction-sc}      % UPO FSSC: (76,6.30)

% 5G construction
\plotpoint{13}{fssc}                 % 5G FSSC: (79,6.30)
\plotpoint{14}{scl8}                 % 5G Fast-SSCL-SPC-8: (116,5.85)
\plotpoint{15}{5g-scl8-crc}              % 5G Fast-SSCL-SPC-8-CRC11: (100,4.92)

% UPO/symmetric construction
\plotpoint{16}{construction-scl8}    % UPO Fast-SSCL-SPC-8: (121,5.45)

% Proposed ensemble decoders
\plotpoint{17}{spae8}                % AE-DISC-8: (40,6.03)
\plotpoint{18}{spae16}               % AE-DISC-16: (40,5.61)
\plotpoint{19}{spae32}               % AE-DISC-32: (40,5.45)
\plotpoint{20}{aed8}                 % AE-SC-8: (77,5.37)

% =========================================================
% Polar (256,128)
% Rows 21--29
% =========================================================
\nextgroupplot[
    xmin=0,
    xmax=190,
    ymin=4.0,
    ymax=6.2,
    ylabel={},
    title={$\mathcal{P}(256,128)$},
    legend to name=decoderlegend,
    legend columns=4,
    legend cell align={left},
    legend style={
        draw=black,
        fill=none,
        font=\small,
        column sep=1.0em,
        nodes={
            minimum width=0pt,
            inner xsep=1pt,
            text depth=0pt
        }
    }
]

% UPO/symmetric construction
\plotpoint{21}{construction-sc}      % UPO FSSC: (77,5.95)

% 5G construction
\plotpoint{22}{fssc}                 % 5G FSSC: (104,5.95)
\plotpoint{23}{scl8}                 % 5G Fast-SSCL-SPC-8: (164,5.55)
\plotpoint{24}{5g-scl8-crc}              % 5G Fast-SSCL-SPC-8-CRC11: (180,4.12)

% UPO/symmetric construction
\plotpoint{25}{construction-scl8}    % UPO Fast-SSCL-SPC-8: (148,4.50)

% Proposed ensemble decoders
\plotpoint{26}{spae8}                % AE-DISC-8: (40,5.22)
\plotpoint{27}{spae16}               % AE-DISC-16: (40,4.90)
\plotpoint{28}{spae32}               % AE-DISC-32: (40,4.59)
\plotpoint{29}{aed8}                 % AE-SC-8: (78,4.39)

% =========================================================
% Legend images (referenced by the custom legend below)
% =========================================================
\addlegendimage{construction-sc,only marks,mark size=2.8pt}
\label{leg:sym-fssc}

\addlegendimage{construction-scl8,only marks,mark size=2.8pt}
\label{leg:sym-fsscl}

\addlegendimage{aed8,only marks,mark size=2.8pt}
\label{leg:ae-sc-8}

\addlegendimage{spae8,only marks,mark size=2.8pt}
\label{leg:ae-disc-8}

\addlegendimage{spae16,only marks,mark size=2.8pt}
\label{leg:ae-disc-16}

\addlegendimage{spae32,only marks,mark size=2.8pt}
\label{leg:ae-disc-32}

\addlegendimage{fssc,only marks,mark size=2.8pt}
\label{leg:5g-fssc}

\addlegendimage{scl8,only marks,mark size=2.8pt}
\label{leg:5g-fsscl}

\addlegendimage{5g-scl8-crc,only marks,mark size=2.8pt}
\label{leg:5g-fsscl-crc}

\end{groupplot}

% =========================================================
% Shared custom legend
% =========================================================
\providecommand{\LatSNRlegend}{%
    \begingroup
    \footnotesize
    \setlength{\fboxsep}{4pt}%
    \setlength{\tabcolsep}{5pt}%
    \renewcommand{\arraystretch}{1.25}%
    \fbox{%
        \begin{tabular}{@{}l@{\hspace{6pt}}c@{\hspace{6pt}}l@{}}

            % Left labels
            \begin{tabular}[c]{@{}l@{}}
                Symmetric Polar / RM \\
                5G Polar
            \end{tabular}

            &

            % Centered vertical separator
            \rule[-2.4ex]{0.4pt}{5.5ex}

            &

            % Legend entries
            \begin{tabular}[c]{@{}llllll@{}}
                \ref{leg:sym-fssc}~FSSC
                    & \ref{leg:sym-fsscl}~FSSCL-8
                    & \ref{leg:ae-sc-8}~AE-SC-8
                    & \ref{leg:ae-disc-8}~AE-DISC-8
                    & \ref{leg:ae-disc-16}~AE-DISC-16
                    & \ref{leg:ae-disc-32}~AE-DISC-32
                    \\
                \ref{leg:5g-fssc}~FSSC
                    & \ref{leg:5g-fsscl}~FSSCL-8
                    & \ref{leg:5g-fsscl-crc}~FSSCL-8-CRC11
                    & & &
            \end{tabular}

        \end{tabular}%
    }%
    \endgroup
}

% =========================================================
% Shared legend below all four plots
% =========================================================
\path
    (group c1r1.south west)
    --
    coordinate (plotcenter)
    (group c4r1.south east);

\node[
    anchor=north,
    inner sep=0pt,
    yshift=-2mm
]
at (plotcenter |- current bounding box.south)
{\LatSNRlegend};

\end{tikzpicture}}
        \caption{
            Required $E_\mathrm{b}/N_0$ at $\mathrm{FER}=10^{-6}$ versus
            decoding latency for decoders.
        }
        \label{fig:rm-latency-reliability}
\end{figure*}

\Cref{fig:rm-latency-reliability} compares the decoding latency and required $E_\mathrm{b}/N_0$ at an \ac{FER} of $10^{-6}$ for the considered RM and Polar decoders. Points closer to the lower-left corner represent more favorable latency--reliability balance. For all four code constructions, \ac{AE-DISC} reduces the decoding latency relative to the corresponding \ac{FSSC} baseline. Increasing the ensemble size further lowers the required \(E_\mathrm{b}/N_0\), progressively shifting the operating point toward the preferred lower-left region. Since all ensemble paths are executed in parallel, increasing the \ac{AE-DISC} ensemble size improves the error-correction performance without increasing the decoding latency. The number of decoding time steps for \ac{FSSCL} is calculated using the model in \cite{fast_scl,doan2022sp}. The fast nodes decoding time steps for \ac{SCL} with list size $L$ are summarized in \Cref{tab:scl_latency}.

For the RM codes, AE-DISC-32 approaches the performance of \ac{AE-SC}-8 with an
$E_\mathrm{b}/N_0$ gap of approximately $0.2$\,dB, while reducing the decoding
latency by $40\%$ and $50\%$ for $\mathrm{RM}(3,7)$ and
$\mathrm{RM}(4,8)$, respectively. A similar trend is observed for the Polar codes. AE-DISC-32 operates within approximately $0.1$\,dB and $0.2$\,dB of \ac{AE-SC}-8 for $\mathcal{P}(128,64)$ and $\mathcal{P}(256,128)$. These results show 
that \ac{AE-DISC} preserves most of the reliability benefit of \ac{AE-SC} while
requiring substantially fewer decoding cycles, demonstrating a favorable
latency-reliability balance for latency-constrained applications.

\section{Hardware Implementation Result}
\begin{figure*}[!t]
    \centering
    \begin{minipage}[t]{0.49\textwidth}
        \centering
        \includegraphics[height=4.2cm]{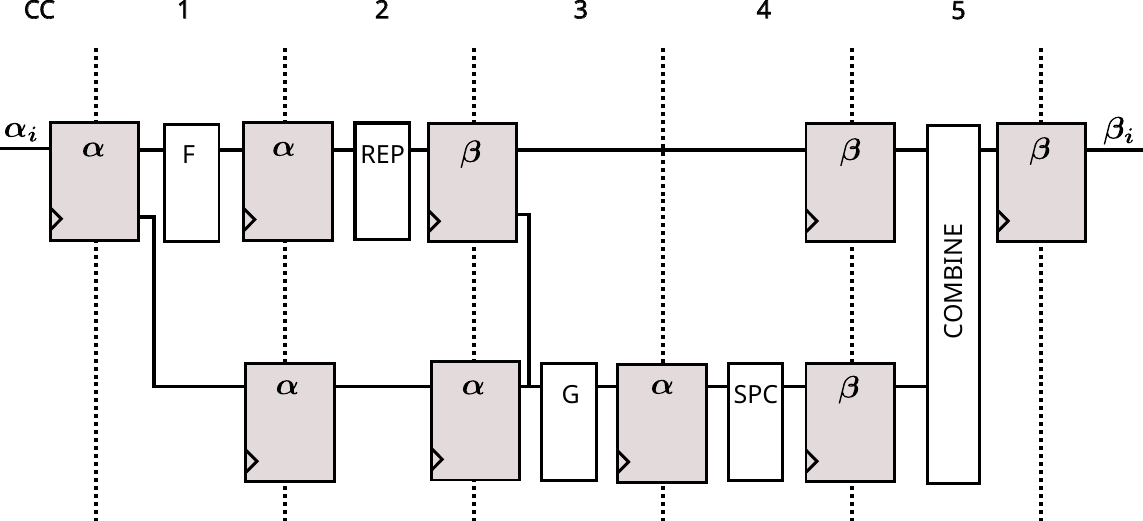}
    \end{minipage}
    \hfill
    \begin{minipage}[t]{0.49\textwidth}
        \centering
        \includegraphics[height=4.2cm]{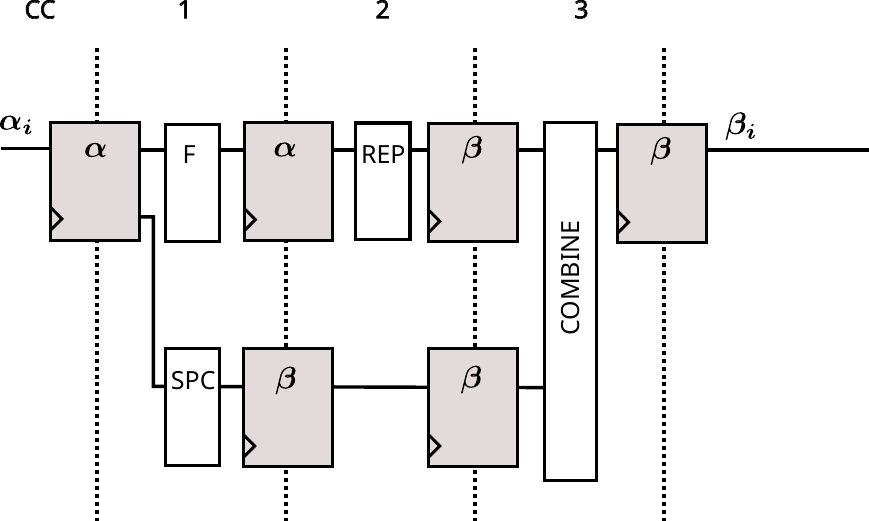}
    \end{minipage}

    \caption{Comparison of architectures for $\mathcal{P}(8,4)$ / $\mathrm{RM}(1,3)$: conventional \ac{FSSC} decoding (left) and the proposed parallel \ac{DISC} decoding architecture (right). }
    \label{fig:unrolled_arch}
    
\end{figure*}

The constituent decoders of the proposed \ac{AE-DISC} decoder are implemented using an unrolled architecture as introduced in \cite{polarbear}. An example of the unrolled architecture for \(\mathrm{RM}(1,3)\) with $N=8$ and $K=4$ is shown in \cref{fig:unrolled_arch}. 
The proposed constituent \ac{DISC} decoder is shown on the right. By executing the left and right branches in parallel, the required decoding latency can be reduced from 5 clock cycles to 3 clock cycles.
The processing blocks in this architecture include REP, SPC, F, G, F0 (an F operation whose left child is an R0 node), and G0 (a G operation whose left child is an R0 node). 

\begin{table*}[!t]
\centering
\caption{\ac{ASIC} implementation results and comparison with other decoder designs.}
\scriptsize
\setlength{\tabcolsep}{3pt}
\def\arraystretch{1.35}
\resizebox{0.9\textwidth}{!}{%
\begin{tabular}{l|ccc|cccc}
\hline
& \multicolumn{3}{c|}{Our Work}
& \multicolumn{4}{c}{Kestel \cite{kestel2023automorphism}} \\
\hline

Design stage
& \multicolumn{3}{c|}{Synthesis}
& \multicolumn{4}{c}{Post-P\&R} \\

Code parameters $(N,K)$
& \multicolumn{3}{c|}{ $(128,64)$}
& \multicolumn{4}{c}{ $(128,60)$} \\

Code rate $R$
& \multicolumn{3}{c|}{ $ 0.5 $}
& \multicolumn{4}{c}{ $ 0.47 $} \\

Decoding algorithm 
& AE-DISC-8
& AE-DISC-32
& AE-SC-8
& AE-SC-8
& SCL-2
& SCL-4
& SCL-8 \\
\hline

Technology $L$ (nm)
& 65
& 65
& 65
& 12
& 12
& 12
& 12 \\

Supply voltage $V_s$ (V)
& 0.9
& 0.9
& 0.9
& 0.9
& 0.9
& 0.9
& 0.9 \\

Area (mm$^2$)
& 1.66
& 6.12
& 2.91
& 0.17
& 0.064
& 0.154
& 0.450 \\

$f_{\max}$ (MHz)
& 263
& 263
& 263
& 498
& 476
& 462
& 478 \\

Throughput @ $f_{\max}$ (Gb/s)
& 33.7
& 33.7
& 33.7
& 63.7
& 61.0
& 59.1
& 61.2 \\

Latency (ns)
& 87.4
& 87.4
& 167.2
& 22.1
& 44.1
& 58.5
& 81.5 \\

Area efficiency (Gb/s/mm$^2$)
& 20.3
& 5.51
& 11.58
& 375.1
& 956.0
& 385.0
& 136.2 \\

\hline

Normalized area$^{a}$ (mm$^2$)
& 1.66
& 6.12
& 2.91
& 4.99
& 1.88
& 4.52
& 13.20 \\

Normalized area efficiency$^{b}$ (Gb/s/mm$^2$)
& 20.30
& 5.51
& 11.58
& 2.36
& 6.02
& 2.42
& 0.86 \\

Normalized throughput$^{c}$ @ $f_{\max}$ (Gb/s)
& 33.70
& 33.70
& 33.70
& 11.76
& 11.26
& 10.91
& 11.30 \\

Normalized latency$^{d}$ (ns)
& \textbf{87.40}
& \textbf{87.40}
& 167.20
& 119.71
& 238.88
& 316.88
& 441.46 \\

\hline
\end{tabular}
}
\par\vspace{5pt} %
{\scriptsize
\begin{tabular}{@{}ll@{}}
Technology scaling with $S=L/{65\,\mathrm{nm}}$: 
&$^{a}\,S^{-2}$, \quad 
$^{b}\,S^3$, \quad
  $^{c}\,S$, \quad
  $^{d}\,S^{-1}$.
\end{tabular}
}
\label{tab:hardware_implementation}
\end{table*}

We implemented the proposed \ac{AE-DISC}-8 and \ac{AE-DISC}-32 decoders for
$\mathcal{P}(128,64)$ using a 65\,nm TSMC technology. For the synthesis stage, Synopsys Design Compiler is used. 
For direct
comparison, an \ac{AE-SC}-8 decoder is also implemented using the same 65\,nm technology
library as the \ac{AE-DISC} decoders. To further ensure a fair comparison with prior works
implemented in different technologies, technology scaling is applied to normalize all
results to a 65\,nm process.

The hardware implementation results are summarized in \cref{tab:hardware_implementation}. 
Both \ac{AE-DISC}-8 and \ac{AE-DISC}-32
achieve substantially lower latency than the \ac{AE-SC}-8 decoder implemented in the same
65\,nm technology. Specifically, the decoding latency is reduced from 167.2\,ns for
\ac{AE-SC}-8 to 87.4\,ns for both \ac{AE-DISC}-8 and \ac{AE-DISC}-32, corresponding to a
47.7\% reduction, closely matching the result from \Cref{ssec:latency}.
This latency reduction is mainly enabled by the proposed parallel
execution of the left and right decoding branches, where the amount of overlap depends
on the node structure. 

Compared with the \ac{SCL} implementations reported by
Kestel~\cite{kestel2023automorphism}, \ac{AE-DISC}-32 also exhibits a clear latency
advantage after normalization to 65\,nm. Specifically, \ac{AE-DISC}-32 achieves a
normalized latency of 87.4\,ns, compared with 238.9\,ns, 316.9\,ns, and 441.5\,ns for
SCL-2, SCL-4, and SCL-8, respectively. Meanwhile, its normalized information
throughput of 33.7\,Gb/s is approximately three times that of the corresponding
\ac{SCL} implementations, which achieve 11.26\,Gb/s, 10.91\,Gb/s, and 11.30\,Gb/s,
respectively. In addition, \ac{AE-DISC}-32 achieves a normalized area efficiency of
5.51\,Gb/s/mm$^2$, outperforming SCL-4 and SCL-8 while providing substantially lower
latency.

More importantly, \ac{AE-DISC}-8 achieves the highest normalized area efficiency among
all compared designs, reaching 20.3\,Gb/s/mm$^2$. Although \ac{AE-DISC}-32 requires a
larger hardware area due to its increased ensemble size, it preserves the same latency
and normalized information throughput as \ac{AE-DISC}-8 because the additional ensemble
paths are executed in parallel. Consequently, increasing the ensemble size from 8 to
32 improves the error-correction performance without introducing additional decoding
latency, at the cost of reduced area efficiency. \ac{AE-DISC} decoders
achieve the highest normalized information throughput and lowest latency among the compared
implementations, highlighting the capability of the proposed architecture for
high-throughput decoding under stringent latency constraints.

\section{Future Works}

Several directions remain for future work. A more comprehensive characterization of the structural properties of the proposed constituent decoder may provide further insights into its applicability to a broader class of codes and decoding frameworks. Also, joint design of code and decoder would provide a promising direction for further improving the error-correction performance and implementation efficiency of the proposed framework. Finally, theoretical analysis of the candidate selection metrics may provide a stronger analytical foundation for the proposed approach and guide future improvements in decoding performance.

\section{Conclusion}
This paper presented \ac{AE-DISC} decoder, a low-latency framework that bridges the gap between reliability and computational efficiency for future high-reliability and high-performance communication systems with Polar-like codes. By enabling concurrent decoding of left and right branches within the \ac{SC} algorithm and exploiting permutation automorphisms, \ac{AE-DISC} significantly reduces decoding latency while preserving state-of-the-art error-correction performance. Simulation results demonstrate that, for the $\mathcal{P}(256,128)$ Polar code, AE-DISC attains up to $1.4$\,dB gain over the 5G \ac{FSSC} baseline and outperforms the five-iteration \ac{LMS} \ac{LDPC} decoder at a \ac{FER} of $10^{-6}$, while maintaining a hardware-friendly complexity. Complexity and latency analysis showed that \ac{AE-DISC} achieves up to $2\times$ speed-up compared to \ac{FSSC} and \ac{AE-SC} decoding. Hardware implementation confirms this latency reduction relative to the \ac{AE-SC} decoder. These results highlight \ac{AE-DISC} as a promising candidate for \ac{HRLLC} in future communication systems, where ultra-low latency decoding architectures are essential.


\vspace{12pt}

\end{document}